\documentclass[compsoc,conference,a4paper,10pt,times]{IEEEtran}
\IEEEoverridecommandlockouts

\usepackage{cite}
\usepackage{ulem}
\usepackage{soul}
\usepackage{booktabs}
\usepackage{amsmath,amssymb,amsfonts}
\usepackage{algorithmicx}
\usepackage{graphicx}
\usepackage{textcomp}
\usepackage{xcolor}
\usepackage{todonotes}
\usepackage[noend]{algpseudocode}
\usepackage{dashbox}
\usepackage[utf8]{inputenc}
\usepackage{tikz}
\usetikzlibrary{circuits.logic.US, circuits.ee.IEC, positioning}
\usepackage[siunitx]{circuitikz}
\usepackage{tcolorbox}
\usepackage{subcaption} 
\usepackage{threeparttable}
\usepackage{pifont}
\usepackage{algorithm}
\usepackage{algpseudocode}
\usepackage{amsmath}
\usepackage{multirow}
\usepackage{tabularx}
 \usepackage{hyperref} 
\usepackage{bbding}
\usepackage{eso-pic}
\newcommand{\zilong}[1]{\textcolor{black}{#1}}
\newcommand{\note}[1]{\textcolor{black}{#1}}
\newcommand{\zl}[1]{\textcolor{black}{#1}}

\usepackage{pgfplots}
\pgfplotsset{compat=1.17}

\usepackage [
    n, 
    advantage,
    operators,
    sets,
    adversary,
    landau,
    probability,
    notions,
    logic,
    ff, 
    mm,
    primitives,
    events,
    complexity,
    oracles,
    asymptotics,
    keys
]{cryptocode}

\newtcolorbox{notebox}{
    colback=red!5!white,
    colframe=red!60!black,
    boxrule=1pt,
    arc=4pt,
    left=6pt,
    right=6pt,
    top=6pt,
    bottom=6pt,
    fontupper=\itshape,
    title=Takeaway:
}

\usepackage{enumitem}
\newcommand*\circled[1]{\tikz[baseline=(char.base)]{%
            \node[shape=circle,fill=blue!20,draw,inner sep=2pt] (char) {#1};}}

\def\BibTeX{{\rm B\kern-.05em{\sc i\kern-.025em b}\kern-.08em
    T\kern-.1667em\lower.7ex\hbox{E}\kern-.125emX}}
\begin{document}
\date{}
\AddToShipoutPictureFG*{%
  \AtPageUpperLeft{%
    \makebox[\paperwidth]{%
      \hspace*{0pt}%
      \raisebox{-1.2cm}[0pt][0pt]{ This paper has been accepted to appear in the IEEE European Symposium on Security and Privacy (Euro S\&P), 2026.}%
    }%
  }%
}
\title{Quantifying Memory Cell Vulnerability for DRAM Security%
\thanks{\Envelope\ ~Corresponding author}
}

\author{

Zilong Hu$^{*}$,
Hongming Fei$^{*}$\Envelope,
Prosanta Gope$^{\dagger}$,
Jack Miskelly$^{\ddagger}$,
Owen Millwood$^{\S}$,
Biplab Sikdar$^{*}$\\[0.5em]

$^{*}$National University of Singapore, Singapore\\
\{huzilong, bsikdar\}@nus.edu.sg, fei.hongming@u.nus.edu\\[0.3em]

$^{\dagger}$University of Sheffield, United Kingdom\\
p.gope@sheffield.ac.uk\\[0.3em]

$^{\ddagger}$Queen's University Belfast, United Kingdom\\
J.Miskelly@qub.ac.uk\\[0.3em]

$^{\S}$Wiznet, Germany\\
owen.millwood@wiznet.eu
}

\maketitle


\begin{abstract}
Dynamic Random Access Memory (DRAM) is pervasive in computer systems. Cell vulnerabilities caused by unintended phenomena (forced retention failure, latency alteration, Rowhammer and Rowpress) lead to unintended bit flips in memory. These phenomena have been explored as attacks to violate data integrity and confidentiality during normal operation, but also exploited as a benefit in security systems as a method to generate random secret keys and unique device fingerprints (e.g. Physically Unclonable Functions). In both cases, attackers may wish to exploit knowledge of individual cell flip vulnerability to predict the current/future data contents of a set of cells, which can be utilised to break security systems.
In this work, we develop a \textbf{quantitative}, \textbf{cell-level} circuit framework that models DRAM vulnerability directly from its physical charge leakage and disturbance pathways. By linking these \textbf{device-layer behaviours} to \textbf{system-level security properties}, our framework enables systematic evaluation of DRAM with respect to \textbf{volatility} (retention), \textbf{integrity} (disturbance-induced modification), and \textbf{confidentiality} (pattern-dependent leakage).
We further demonstrate how the framework can be applied to well-known failure modes, revealing non-uniform and context-dependent vulnerability patterns. 
This work provides both \emph{theoretical foundations} and \emph{practical} evaluation tools for evaluating the suitability of DRAM use within security applications.

\end{abstract}

\section{Introduction}
\label{sec1}
When it comes to these cell-level effects, two mechanisms are particularly influential. First, \emph{retention loss} describes the natural decay of stored charge in a DRAM capacitor; without timely refresh, the storage node voltage drifts toward the sense threshold, eventually leading to a bit flip. Second, \emph{read disturbance}---popularly known as \zilong{Rowhammer and Rowpress} \cite{kim2014flipping,luo2023rowpress}---describes the effect where repeated or sustainable activations of one or more rows can induce bit flips in neighbouring rows even in the absence of write permissions. \zilong{Both mechanisms have been observed in real systems and weaponized in attacks~\cite{lin2025gpuhammer,cojocar2019exploiting,kwong2020rambleed}, however, both the security community and DRAM vendors mostly treat these failures empirically: attacks are tuned to device-specific behaviours, and mitigations are evaluated by ad-hoc testing. A comprehensive understanding of the cause is lacking, largely due to the absence of an \textbf{attacker-centric, quantitative model} for DRAM that links measurable cell-level parameters to concrete security properties (integrity, confidentiality, volatility).}

\zilong{To fulfill the gap and give a comprehensive analysis, we proposed a security-oriented framework: it aggregates physical leakage mechanisms into a small set of attacker-interpretable parameters ($R_S$, $R_B$, $N$), and connects these parameters to attack objectives such as precise bit flip control and pattern-based key inference to reveal the essence of the DRAM attacks from a theoretical to empirical way.
This lets us reason about which cells and patterns are exploitable, rather than just how often errors occur.}
\zilong{Moreover, we evaluate the \textbf{existing} attack surfaces systematically, then provide insights into potential attack surfaces.}




\subsection{Related Work}
\zilong{In recent years, the vulnerability of DRAM cells has been explored through empirical studies.
Liu et al. \cite{liu2013experimental} showed that data patterns significantly affect retention-induced bitflips due to parasitic coupling between adjacent wordlines and bitlines.
Building on this, the Rowhammer effect \cite{kim2014flipping} demonstrated that repeated activation of aggressor rows can externally amplify cell fragility through disturbance-induced charge leakage, where the number of bitflips also depends on the aggressor–victim data patterns. \emph{CheckerBoard} (i.e., alternating $010$ between columns) is usually the most efficient pattern, while \emph{Stripe} (same value for the entire row, such as a consistent $111$ pattern) becomes weaker.
Yağlıkçı et al. \cite{yauglikcci2022understanding} demonstrate that tampering with wordline voltages can alter the rowhammer behaviour and increase vulnerability.
Orosa et al. \cite{orosa2021deeper} demonstrate that the Rowhammer effect is more likely to occur when aggressor rows remain active for a longer duration.
Furthermore, the RowPress effect \cite{luo2023rowpress} revealed that even static but prolonged wordline activation can amplify the vulnerability of the memory cells. 
Luo et al. \cite{luo2025revisiting} also reported inconsistency between previously proposed error mechanisms and pattern dependency effects with empirical results from 96 COTS DDR4 DRAM chips. 
Together, these mechanisms reveal that DRAM cell fragility arises not only from intrinsic retention loss but also from pattern-dependent disturbance and field-induced leakage for which modeling capabilities are currently limited.}
\zilong{While existing studies provide valuable empirical observations on cells vulnerablity, many researchers \cite{halderman2009lest,lin2025gpuhammer,cojocar2019exploiting,kwong2020rambleed} have also made use of these observations to perform real attacks. The most classic example is the Cold Boot Attack\cite{halderman2009lest} which exploits the fact that DRAM, while nominally volatile, has a certain time before actual physical retention failure. That time is highly dependent on temperature, so rapidly cooling the DRAM after power off allows partial reconstruction of sensitive data such as disk encryption keys once power is restored. This is an example of a \emph{volatility} attack.}
Lin et al. \cite{lin2025gpuhammer} show that Rowhammer extends to discrete GPUs with GDDR6, where induced bit-flips can severely degrade machine learning model \emph{integrity}, underscoring its system-level security impact. Lucian et al. \cite{cojocar2019exploiting} show that Rowhammer can be used to flip the targeted bits and keep the other bits unchanged by controlling the access patterns, another integrity attack. \zilong{In addition to this kind of integrity violation, Kwong et al. \cite{kwong2020rambleed} further show that Rowhammer can be used in a \emph{confidentiality} attack by observing flip patterns in attacker-controlled memory to infer victim data (e.g., cryptographic keys). These physical level effects touch on all security properties of DRAM.}

 Jiang et al. \cite{jiang2021quantifying} attempted to quantify Rowhammer at the circuit level by modelling the leakage mechanism. This study shows that when a victim cell’s storage resistance ($R_{SW}$) and leakage resistance ($R_{L}$) are fixed, the activation time of the adjacent aggressor row directly determines the likelihood of disturbance-induced bitflips. This provides a circuit-parameter-centric perspective on how Rowhammer arises from cell-level electrical constraints. On the other hand, Walker et al. \cite{walker2021dram} derive a mathematical and physics-based formulation of Rowhammer, offering a more analytical expression for how repeated activations couple into victim cells to induce charge leakage. Together, these two works illustrate complementary approaches: while Jiang et al. \cite{jiang2021quantifying} emphasize empirical quantification through circuit parameters, Walker et al. \cite{walker2021dram} highlight analytical modelling grounded in physical equations. Both perspectives converge to deepen our understanding of Rowhammer as a phenomenon governed by the joint properties of electrical devices and activation-induced disturbance physics.

\subsection{Motivation}
\label{sec:motivation}

\zilong{In summary, prior work has either (i) empirically demonstrated that DRAM failures can be weaponized in a number of attack surfaces, or (ii) built physical models to explain these failures, mostly for reliability purposes.
To the best of our knowledge, no prior work has combined a cell-level physical model with a systematic, attacker-centric security evaluation, nor demonstrated that the underlying parameters can be reliably measured on commodity devices and used to compare the impact of different failure mechanisms on attack surfaces.
This is precisely the gap our work aims to address.}

\begin{figure}[t]
  \centering
  \includegraphics[width=0.95\linewidth]{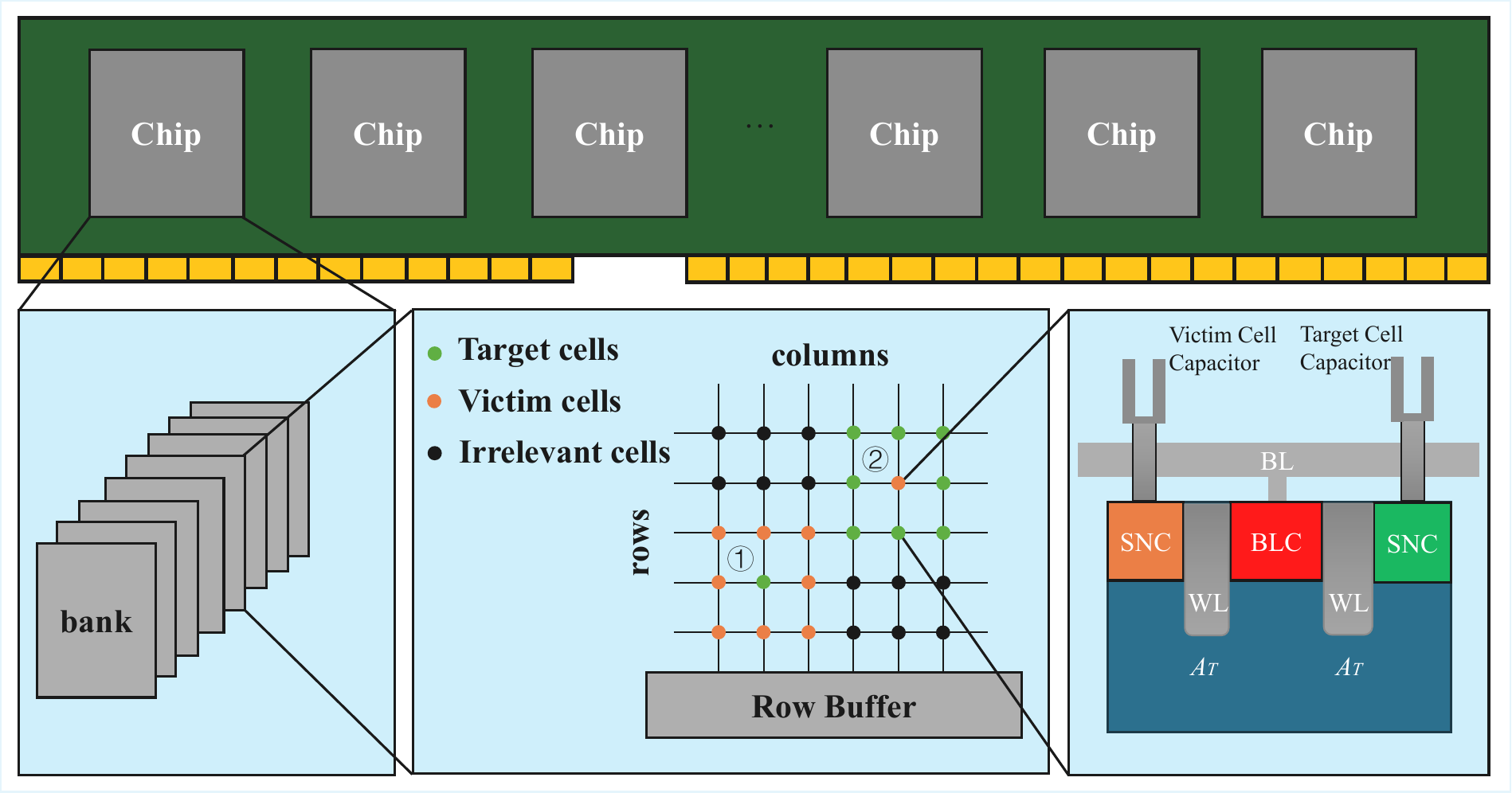}
  \caption{An overview of DRAM organization. Independent cell capacitors share substrate, BL, and WL, inducing pattern-dependent coupling. Scenario \textbf{\ding{192}} and \textbf{\ding{193}} represent two attack case studies as discussed in Section~\ref{sec4:memory_model}.}
  \label{fig:dram_org}
\end{figure}

To bridge this gap, we focus on a key phenomenon that naturally links physical cell behaviour with attacker capabilities: \textbf{pattern awareness}.
Although adjacent DRAM cells store charge in independent capacitors (Figure~\ref{fig:dram_org}), they are not isolated from each other.
They share the same p-type substrate, bitlines (BL), and wordlines (WL), which introduce capacitive and subthreshold coupling.
As a result, a cell’s flip probability is determined not only by its own device parameters and intrinsic leakage, but also by the \emph{surrounding data pattern} encoded in its neighborhood.

From an attacker’s perspective, this pattern awareness immediately gives rise to two distinct classes of attack surfaces:
(i) \emph{pattern control}, where the adversary deliberately \emph{sculpts} neighbour data so as to amplify or suppress a victim cell’s flip probability (targeted integrity violation); and
(ii) \emph{pattern inference}, where the adversary \emph{measures} flip statistics on carefully chosen probe cells in order to infer unknown neighbor data (confidentiality violation).
These behaviors have been repeatedly observed in empirical studies, but in the absence of a quantitative, cell-level model, it remains unclear how to systematically evaluate which cells, patterns, and devices are exploitable, or how different failure mechanisms (e.g., retention, Rowhammer, Rowpress) compare on the same footing.

\zilong{This leads to a broad central question:
From the attacker's perspective, can we turn measurable cell-level DRAM parameters into quantitative indicators of how exploitable a given device is for volatility-, integrity-, and confidentiality-based attacks?}
More specifically, we aim to answer the following research questions:


\begin{enumerate}
    \item[\textbf{RQ1:}] How can we formalize a physical DRAM instance as \zilong{an explainable model} with well-defined, measurable parameters \zilong{\emph{from the attackers' perspective}}?
        \item[\textbf{RQ2:}] \zilong{ Can our model be exploited to accurately qualify or evaluate the performance of integrity and confidentiality attacks on the target device? }
    \item[\textbf{RQ3:}] If so, can the model provide new insights to existing attack surfaces not seen in empirical analysis? 
\end{enumerate}



\subsection{Our Contributions}
To answer the above research questions and bridge the gap between empirical DRAM characterization and \emph{formal} security evaluation, we present a \textbf{model-driven, cell-level framework} that explains and quantifies DRAM physical behaviours from a security perspective. In Section~\ref{sec2}, we introduce the background knowledge. After illustrating the threat model in Section~\ref{sec3:threat_model}, we present the proposed memory circuit model, two case studies and detailed attack qualification in Section~\ref{sec4:memory_model} and make the following key contributions:

\begin{enumerate}[label=\protect\circled{\arabic*}]
\setlength\itemsep{-0.25em}
    \item \zilong{To answer the \textbf{RQ1}, }We develop the \textit{first} empirically verified cell circuit-level model that jointly characterizes DRAM volatility and read disturbance to \emph{qualify} attacks from a \emph{security} perspective. This model yields closed-form conditions for when a given cell is exploitable for  \textit{arbitrary and targeted bit flips} or \textit{pattern inference}, as a function of $R_S$ and $R_B$ . 
    \item \zilong{To answer the \textbf{RQ2}, we collect measurable electrical parameters on our FPGA-based DRAM testing platform (Xilinx Alveo U200 + DRAM Bender),
    and translate them into security-relevant objectives
    . We show that these parameters can be robustly estimated on } \note{seven} commodity DDR4 modules, and we characterize how exploitable vulnerabilities impact existing and potential attack surfaces.
    \item \zilong{To answer the \textbf{RQ3}, using our model-driven framework, we uncover potential RowPress-based attack surfaces. Our evaluation reveals, for instance, that RowPress offers significantly better selectivity for \textit{targeted bit flip} and \textit{pattern-based inference} than conventional Rowhammer on the tested devices, as demonstrated through theoretical analysis and chip-level experiments, suggesting new attack vectors and directions for mitigation.
    }

\end{enumerate}

The remainder of this paper is organized as follows. 
Section~\ref{sec2} provides background on DRAM architecture and summarizes known cell-level vulnerabilities that motivate our study. 
Section~\ref{sec3:threat_model} introduces the threat model, defining the security properties of DRAM and outlining adversarial capabilities in common application scenarios. 
Section~\ref{sec4:memory_model} presents our cell-level circuit model, which formalizes the physical mechanisms underlying retention and disturbance failures and connects them to concrete case studies and attack qualification. 
Section~\ref{sec5:evaluation} reports our experimental methodology and results on commodity DRAM devices, and demonstrates how the proposed parameters can be measured and adopted into the attack surfaces. 
Finally, Section~\ref{sec6:conclusion} concludes the paper with a discussion of broader implications.



\section{Preliminaries}
\label{sec2}


In this section, we first introduce the two classic disturbance-based attacks, the Rowhammer attack and the Rowpress attack.
We then describe the underlying DRAM cell vulnerabilities that enable these attacks and the resulting attack surfaces.
Finally, we discuss the threat model and define the adversary capabilities used in our security evaluation.

\subsection{DRAM Cell Vulnerabilities and Attack Surfaces}
\label{sec:DRAM Cell Vulnerabilities}

\textbf{Rowhammer Attack.}
Rowhammer is a disturbance attack in which an adversary repeatedly activates and precharges one or more DRAM rows (the \emph{aggressor rows}) at a high rate, causing charge to leak from physically adjacent \emph{victim rows}.
When the cumulative disturbance is large enough, victim cells experience accelerated charge loss and may flip their stored bits without any explicit write to the corresponding memory region, leading to a violation of memory integrity.
Over the past decade, Rowhammer has been shown to corrupt page tables, escalate privileges, and compromise isolation boundaries on a wide range of commodity systems~\cite{halderman2009lest,cojocar2019exploiting}.
In all cases, the attack fundamentally exploits how repeated activations modulate the victim cell’s effective leakage paths and shorten its time-to-failure.

\textbf{Rowpress Attack.}
Rowpress is a closely related disturbance mechanism in which the adversary does not necessarily toggle the aggressor row at a high frequency, but instead keeps a row \emph{open} (i.e., wordline asserted) for an extended period.
This long-duration activation places sustained electrical stress on cells that share substrate and bitline structures with the pressed row, altering their leakage behaviour.
Compared to Rowhammer, Rowpress can induce bitflips with fewer activation cycles and exhibits a stronger dependence on the stored data pattern in neighbouring rows, making it particularly relevant for \emph{pattern-aware} integrity and confidentiality attacks~\zl{\cite{luo2023rowpress}}.
In this work, we will treat Rowhammer and Rowpress as two distinct, but closely related, read-disturbance mechanisms acting on the same underlying cell-level vulnerabilities.

\zilong{
Existing work has demonstrated numerous feasible attacks that exploit memory cell vulnerabilities~\cite{halderman2009lest,lin2025gpuhammer,cojocar2019exploiting,kwong2020rambleed}. 
Although these attacks are based on real empirical observations of cell behaviour, it remains unclear which physical leakage paths truly matter for each attack surface.
In order to examine this more closely, we must establish the relevant leakage paths, the associated electrical properties, and their relation to the security properties of \emph{volatility}, \emph{integrity}, and \emph{confidentiality}.}

A DRAM cell typically consists of a single access transistor ($A_T$) and a storage capacitor ($C_\mathrm{s}$), forming a 1T1C (one-transistor-one-capacitor) architecture.
As illustrated in Fig.~\ref{fig:DRAM_cell}, one terminal of the storage capacitor is connected to a stable reference voltage, typically $V_\mathrm{DD}/2$ or ground.
In contrast, the other terminal is connected to the drain (or source) of $A_T$, whose source (or drain) is in turn connected to the bitline that interfaces with the sense amplifier. 

\zilong{
We consider two primary leakage paths for the charge maintained in the storage capacitor $C_\mathrm{s}$.
These can be described in terms of two leakage currents, $I_1$ and $I_2$.
The first, $I_1$, is the path from the storage node to \textbf{ground} through the p-well; the second, $I_2$, is the path from the storage node to the \textbf{bitline}.}
\zilong{
Because cells along a column share the same bitline, the $I_2$ path is inherently influenced by neighbouring cells.
This shared structure allows disturbances and side-channel information to propagate through the bitline-coupled leakage, making $I_2$ crucial for understanding pattern-dependent attack surfaces.}

\zilong{
Throughout the rest of this paper, we will use a small set of effective parameters to summarize these two physical leakage paths.
Specifically, we denote by $R_S$ the effective resistance of the p-well leakage path associated with $I_1$, and by $R_B$ the baseline resistance of the bitline-coupled leakage path associated with $I_2$ under a neutral neighbour pattern.
In addition, we introduce a pattern-dependent noise term $N(P)$ to capture how different neighbour patterns $P$ modulate the bitline voltage and thus the effective leakage along the $I_2$ path.
Intuitively, $R_S$ primarily governs volatility and coarse integrity properties, while $R_B$ and $N(P)$ are the key determinants of pattern-dependent integrity and confidentiality attacks.
These quantities $(R_S, R_B, N(P))$ form the core parameters of our cell-level circuit model in Section~\ref{sec4:memory_model} and will be explicitly measured and evaluated on real DRAM chips in Section~\ref{sec5:evaluation}.
}

\subsubsection{P-well Leakage ($I_1$)}

\paragraph{\textbf{Intrinsic Leakage}}
\label{Junction Leakage Path}
Without cell refresh, the charge maintained in $C_\mathrm{s}$ will leak over time and eventually fail.
The charge variation in the upper plate of the storage capacitor $C_\mathrm{s}$ determines the failure characteristics of the cell. 
\zilong{
Two major components determine the intrinsic leakage through $I_1$.
First, transistor leakage, mainly due to Gate-Induced Drain Leakage (GIDL), where a strong electric field near the drain junction under negative gate bias triggers band-to-band tunnelling (BTBT), generating unwanted current~\cite{chen2001analytic}. 
Second, dielectric leakage between the bitline contact (BLC) and storage node contact (SNC), which worsens with technology scaling as the dielectric spacing shrinks.
Intrinsic leakage caused by these two behaviours forms the characteristic \emph{\textbf{volatility}} of DRAM cells and does not require any external disturbance to occur.} 

\paragraph{\textbf{Electron Injection Leakage}}
In \zilong{read-disturbance} scenarios, frequent activations of aggressor rows repeatedly toggle their wordlines, creating transient high electric fields in the substrate.  
\zilong{
These fields inject a small population of electrons into the p-well region beneath adjacent victim cells.
Once injected, these electrons diffuse laterally and vertically, and some reach the victim cell's storage node, effectively accelerating its charge leakage $I_1$.}  

This mechanism differs fundamentally from retention failure, in which leakage is driven solely by passive decay paths, without external disturbance.
Moreover, in contrast to the slow, steady nature of retention-based charge decay, the injection-induced leakage current in Rowhammer scenarios can be significantly larger due to the strong, repetitive electric-field stress.
\zilong{
This electron-injection leakage under Rowhammer disturbance gradually discharges the cell until it flips, and is a primary cause of DRAM \emph{\textbf{integrity}} violations.}

\subsubsection{Bitline Leakage ($I_2$)}
\label{Sec:Bitline Leakage}
The second pathway ($I_2$) involves charge sharing between the bitline and the storage capacitor, mediated by $A_T$.

\paragraph{\textbf{Sensing Noise}}
\label{sec:Sensing Noise}

During the sensing phase of a DRAM read, the bitline voltage changes only slightly, on the order of tens of millivolts, due to capacitive charge sharing with the storage capacitor~\cite{sekiguchi2002low}.  
This small swing must be amplified by the sense amplifier (SA).  
However, in this narrow amplification window, parasitic capacitive coupling from nearby structures can inject unwanted noise onto the bitline, a phenomenon known as \textit{sensing noise}.  

Under retention failure, a key feature of sensing noise is that it can induce \textbf{data-dependent} bitflips~\cite{liu2013experimental}.  
For instance, if a neighbouring bitline stores a \texttt{0}, its falling transition during sensing may couple negatively onto an adjacent bitline storing a \texttt{1}, suppressing its voltage rise.  
Such cross-bitline interference reduces the effective sensing margin and \zilong{reduces the effective retention time of memory cells, thereby \emph{amplifying} DRAM \emph{\textbf{volatility}}.}

\paragraph{\textbf{Subthreshold Leakage}}

\zilong{
As shown in Fig.~\ref{fig:DRAM_cell}, when a victim row is under Rowhammer attack, the row's wordline is partially activated by the crosstalk coupling induced by repeated aggressor row activations.
The voltage between bitline and storage capacitor is $V_\mathrm{BS}$, and this voltage difference leads to a subthreshold leakage current $I_2$ from the storage node to the bitline.
Past a certain threshold of prolonged disturbance, $V_S$ will decrease below $V_\mathrm{DD}/2$ and the cell will flip from $1$ to $0$.
However, the bitline voltage $V_\mathrm{B}$ fluctuation is determined by the values of the cells in the aggressor rows.
If the cells in aggressor rows store a low voltage level, they pull $V_\mathrm{B}$ towards $0$ and increase $V_\mathrm{SB}$, thus amplifying the leakage current $I_2$.
In contrast, if the cells store a high voltage level, they pull $V_\mathrm{B}$ towards $V_\mathrm{DD}$ and decrease $V_\mathrm{SB}$, thereby suppressing the leakage current $I_2$.
This subthreshold leakage, adjusted by the victim and aggressor data patterns, not only amplifies the impact of \emph{\textbf{integrity}} violations, but also serves as a side channel that can be exploited to violate \emph{\textbf{confidentiality}}.}

\begin{figure}[t]
  \centering
  \begin{tikzpicture}[circuit ee IEC, scale=0.8, transform shape,, every node/.style={font=\small}]
      \draw (0.65,-0.5) -- (0.65,3.5);
      \node[anchor=south] at (0.3,3.5) {Bitline};

      \draw (-1.5,2.5) -- (1.5,2.5);
      \node[anchor=south] at (1.5,2.5) {Wordline};

      \draw (0,2.5) node[nmos, anchor=gate, rotate=-90, scale=1] (T1) {};
      \node at ($(T1)+( -0.5, 0.5 )$) {$A_T$};

      \draw[->, thick, blue] ($(T1.source)+(0.3,-0.2)$) -- ++(0.8,0)
          node[midway, above] {$I_\mathrm{2}$};

      \fill (T1.source) circle (1.2pt);
      \fill (T1.gate) circle (1.2pt);
      \node[above = 2pt of T1.gate] {$V_W$};

      \node[below left=1pt and 2pt of T1.source] {$V_S$};
      \fill (T1.drain) circle (1.2pt);
      \node[below right=1pt and 2pt of T1.drain] {$V_B$};

      \draw (T1.source) -- ++(0,-0.5) node[coordinate] (cjoint) {};
      \draw (cjoint) to[C, l_=$C_\mathrm{s}$] ++(0,-1) node[coordinate] (cjoint2) {};

      \draw[->, thick, blue] ($(cjoint2)+(0.5,0.8)$) -- ++(0,-0.6) node[midway, right] {$I_\mathrm{1}$};

      \node[anchor=south] at ($(cjoint2)+(0,-0.5)$) {$V_{\mathrm{CP}}\!\approx\!V_{DD}/2$};
    \end{tikzpicture}
    \caption{DRAM cell leakage paths under retention and read disturbance.}
    \label{fig:DRAM_cell}
\end{figure}

\subsection{Threat Model}
\label{sec3:threat_model}


Before discussing the circuit-level models, we first define the threat model that guides our interpretation of DRAM cell parameters from a security perspective.
Our goal is not to propose a new end-to-end exploit, but to provide a quantitative framework that characterizes how different DRAM devices behave under realistic adversarial capabilities.

\paragraph{\textbf{Security Properties}}
Our analysis is organised around three core properties of memory:

\begin{itemize}
    \item \textbf{Memory Integrity}: values in memory cannot be altered except through intended channels and with valid write access to the relevant memory region.
    \item \textbf{Memory Confidentiality}: values in memory cannot be learned except through intended channels and with valid read access to the relevant memory region. 
    \item \textbf{Memory Volatility}: without active maintenance (refresh), the values in memory are irrecoverably lost. 
\end{itemize}

\paragraph{\textbf{Abstract Adversary Capabilities}}
We consider an \emph{attacker-oriented} viewpoint in which the adversary interacts with DRAM only through architecturally allowed memory operations, but may have advanced control over how these operations are issued.
In particular, we assume the adversary may possess some combination of the following high-level capabilities:
\begin{itemize}
    \item \textbf{Pattern control}: the ability to place chosen data patterns $P$ in selected memory regions (e.g., rows surrounding a victim cell) and to issue repeated accesses such as Rowhammer or Rowpress-style activations.
    \item \textbf{Observation}: the ability to repeatedly probe memory locations and observe whether bitflips occur, or whether previously written data can be partially recovered after a disturbance or power cycle.
    \item \textbf{Timing/refresh control}: in some scenarios, the ability to influence refresh behaviour or power cycles (e.g., cold-boot-like settings), thereby exposing retention-related failures.
\end{itemize}
These abstract capabilities capture a broad range of known DRAM-based attacks (e.g., Rowhammer, Rowpress, cold boot, RAMBleed) without committing to a particular exploit chain.
The detailed mapping to concrete attacker models and system privileges is given in Appendix~C.

\paragraph{\textbf{Adversarial Objectives}}
Given these capabilities, we distinguish between \emph{coarse} objectives, which do not exploit pattern dependence, and \emph{pattern-aware} objectives, which do.

\medskip
\noindent\textbf{Coarse objectives.}
\begin{itemize}
    \item \textbf{OBJ-I-1 (Integrity-1)}: alter \emph{any} bits in a write-protected memory region, i.e., a coarse integrity violation.
    \item \textbf{OBJ-C-1 (Confidentiality-1)}: learn \emph{any} bits in a read-protected memory region, i.e., a coarse confidentiality violation.
    \item \textbf{OBJ-V-1 (Volatility-1)}: partially retrieve bits of memory contents that should have been lost after refresh stops or after a power cycle, i.e., a coarse volatility violation.
\end{itemize}

\noindent\textbf{Pattern-aware objectives.}
Knowledge of pattern dependence and the ability to control patterns allow the adversary to refine these into more targeted objectives:
\begin{itemize}
    \item \textbf{OBJ-I-2 (Pattern-control integrity)}: by writing specific data patterns in cells surrounding a victim cell, increase the victim’s flip probability $p_{\mathrm{flip}}(P)$ and enable \emph{pattern-sculpted} targeted modifications (e.g., steering bitflips to particular locations).
    \item \textbf{OBJ-C-2 (Pattern-inference confidentiality)}: by observing changes in a probe or victim cell’s flip probability across controlled disturbance trials, infer unknown neighbour data (the \emph{surrounding pattern}).
    \item \textbf{OBJ-V-2 (Pattern-regeneration volatility)}: by modelling retention failure characteristics and reading the start-up state of a specific victim cell after power-up, infer probable pre-power-down states of the stored data.
\end{itemize}


In summary, this threat model abstracts away system-specific exploit chains and instead focuses on \emph{which cell-level parameters an adversary can, in principle, leverage to violate integrity, confidentiality, or volatility}.
Our circuit model then provides a quantitative link between these parameters and the attacker objectives above, enabling a model-driven comparison of different DRAM devices and disturbance mechanisms.



\section{Cell-Level Circuit Model and Attack Qualification}
\label{sec4:memory_model}

\begin{figure}[t]
  \centering
  \begin{subfigure}{\linewidth}
    \centering
    \includegraphics[width=0.6\linewidth]{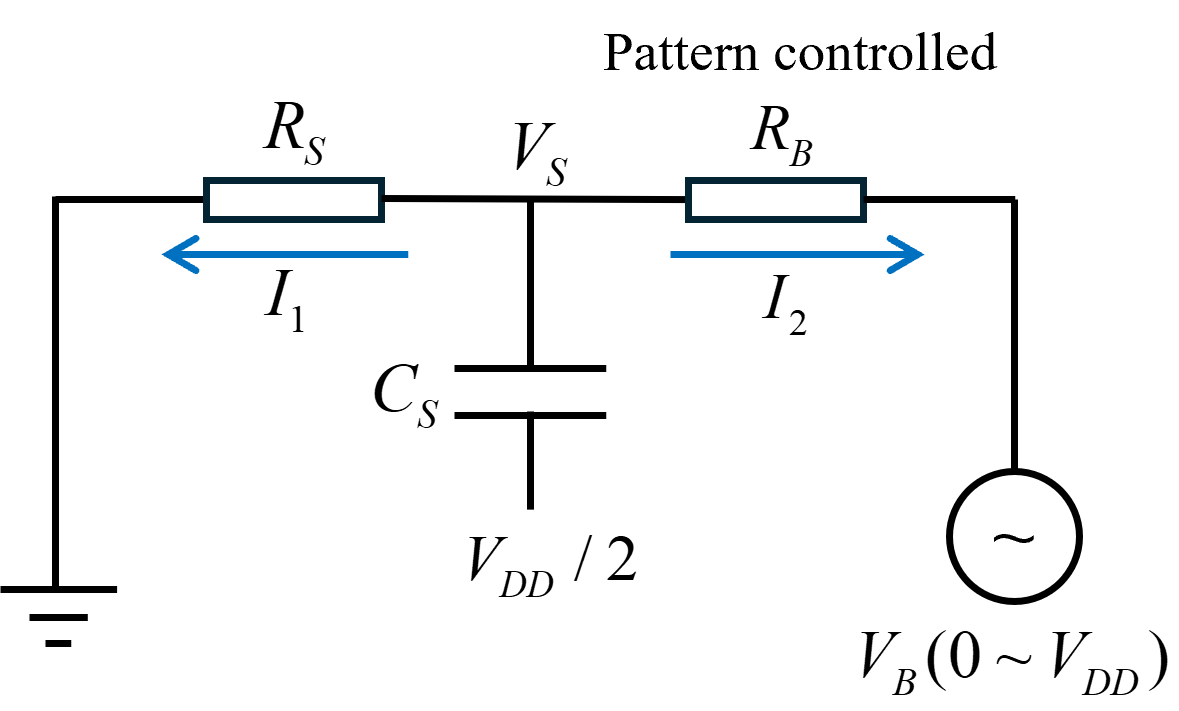}
    \label{fig:rs_dis_b}
  \end{subfigure}
  \caption{Equivalent Leakage Circuit.}
  \label{fig:DRAM_cell_2}
\end{figure}

In this section, we start by presenting a unified cell-level leakage model that explains three classes of DRAM security–relevant behaviors: \emph{volatility} (retention loss), \emph{integrity violation} (disturbance-induced flips), and \emph{confidentiality leakage} (pattern inference). 
\zilong{As illustrated in Fig. \ref{fig:DRAM_cell_2}, our cell-level circuit leakage model abstracts the physical cell circuit into a $RC$ circuit. First, we characterize two primary leakage paths as two resistances $R_S$ and $R_B$. Then, we assume the current $I_1$ leaks through $R_S$ to the ground and $I_2$ leaks through $R_B$ to the bitline. Under attack scenarios, bitline voltage will be modulated from $0$ to $V_{DD}$ due to the disturbance from the repeated read operations. Therefore, $I_2$ will be influenced by the data pattern in adjacent rows through the bitline. 
Using this model as a baseline we can delve into volatility, integrity, and confidentiality situations, providing a precise model of the cell vulnerability for each of them.}

\subsection{Volatility Model}
From the perspective of volatility, we are interested in failures that occur \emph{without any explicit disturbance}, purely due to intrinsic charge leakage.
A DRAM cell is initially charged to $V_{DD}$ (logic \texttt{1}) and then left idle. A bitflip occurs when the held charge decays sufficiently that the wordline voltage is mischaracterized by the sense amplifier during read operations. During sensing, coupling noise from adjacent wordlines and bitlines can further reduce the margin of error, increasing the chance of a bitflip.

Accordingly, in the volatility model we consider only the leakage path $I_1$ through $R_S$ and treat the bitline branch as effectively open (\(R_B^{(\mathrm{eff})}\!\to\!\infty\)).
We then extend this to include pattern-induced sensing noise.

\subsubsection{Noise-free Leakage Model}

Retention failure occurs when the charge stored in the cell capacitor gradually leaks over time, causing the storage voltage to drop below the sensing threshold. \zilong{We first consider all the neighboring cells hold the high voltage level so sensing noise can be neglected.}
During the idle period, the wordline is held low ($V_W = 0$), while the bitline is precharged to $V_{DD}/2$.  
The storage node voltage $V_S$ is initialized to $V_{DD}$ (logic \texttt{1}).  \zilong{Considering only the intrinsic leakage $I_1$ in this setting,}
the behavior can be modeled by an equivalent $RC$ circuit, where the capacitor $C_S$ represents the storage element and the resistor $R_S$ denotes the \textbf{intrinsic leakage resistance}.  

The temporal evolution of the storage voltage follows a first-order $RC$ discharge process:
\begin{equation}
\frac{dV_{S}(t)}{dt} = -\frac{1}{R_{S} C_S}V_S(t)
\quad \Rightarrow \quad
V_S(t) = V_{DD} \cdot e^{-t/(R_{S} C_S)}.
\end{equation}
Here, the time constant $\tau = R_{S} C_S$ characterizes the retention capability of the cell, with reliable sensing typically guaranteed only within \zl{one time constant} $\tau$.  

The instantaneous leakage current is given by
\begin{equation}
I_{1}(t) = C_S \cdot \frac{dV_S(t)}{dt}
= -\frac{V_{DD}}{R_{S}} \cdot e^{-t/(R_{S} C_S)}.
\end{equation}
This expression shows that the leakage current is inversely proportional to $R_{S}$ and decreases exponentially over time.  
Cells with smaller $R_{S}$ discharge more quickly, making them more vulnerable to retention failures.

\zilong{Then the DRAM cell is accessed during a read operation, the stored charge in the capacitor is shared with the precharged bitline.  
This sharing produces a small voltage swing on the bitline given by}

\begin{equation}
\label{vbl}
\Delta V_{\mathrm{BL}} = \frac{C_S}{C_S + C_{\mathrm{BL}}} \cdot \left( \frac{V_{DD}}{2} - V_S(t) \right),
\end{equation}

where $C_{\mathrm{BL}}$ is the bitline capacitance.  
The ratio $\alpha = \tfrac{C_S}{C_S + C_{\mathrm{BL}}}$ characterizes the fraction of the storage charge that is transferred to the bitline during sensing. The resulting signal $\Delta V_{\mathrm{BL}}(t)$ is then amplified by the sense amplifier; \zilong{if $|\Delta V_{\mathrm{BL}}(t)|$ does not exceed the sensing threshold $V_{SA}$, the sense amplifier may misread the cell and the stored data is effectively lost.}



\subsubsection{Pattern-noised Leakage Model}

As discussed in Section~\ref{Sec:Bitline Leakage}, when the surrounding cells are biased to a low-voltage state, they inject additional charge into the shared bitline through bitline and wordline capacitive coupling. 
We model the resulting \emph{pattern-induced noise} as an effective voltage perturbation $N$ on the bitline.
Taking this noise into account, the readout condition becomes

\begin{equation}
\left\{
\begin{aligned}
&\Delta V_{\mathrm{BL}} + N > V_{SA}, && \text{Correct read (still '1')} \\
&\Delta V_{\mathrm{BL}} + N < -V_{SA}, && \text{Flip error (misread as '0')} \\
&|\Delta V_{\mathrm{BL}} + N| < V_{SA}, && \text{Uncertain outcome}
\end{aligned}
\right.
\end{equation}

In pure volatility, the bitline branch is effectively closed, \(R_B^{(\mathrm{eff})}\!\to\!\infty\), so \(\tau \approx C_S R_S\).  
Here integrity errors are time–pattern dependent through \(R_S\) and \(N\): the flip probability increases with longer dwell time and larger \(|N|\), and the attack knob is to \emph{extend refresh interval} or \emph{bias patterns} that reduce the effective \(R_S\) (coupling-assisted leakage).  
Formally, referring to equation \ref{vbl}, a flip occurs when
\begin{equation}
\frac{C_S}{C_S + C_{\mathrm{BL}}} \cdot \left( \frac{V_{DD}}{2} - V_S(0)\,e^{-T/(C_S R_S)}  \right)
+N < -V_{SA}.
\end{equation}
Thus, \zilong{in the pattern-noised leakage model, both the intrinsic leakage resistance $R_S$ and the external coupling noise $N$ jointly determine the cell's data \emph{volatility}.  
Larger noise magnitudes $|N|$ reduce the sensing margin and increase the probability that a cell becomes vulnerable to retention-induced bitflips.}


\subsection{Read Disturbance Model}
\zilong{Read-disturbance phenomena such as Rowhammer and Rowpress introduce failure modes that go beyond pure volatility. 
In the integrity and confidentiality model, we consider both leakage paths, $I_1$ and $I_2$. 
As illustrated in Fig. \ref{fig:DRAM_cell_2}, we customize their leakage resistances $R_S$ and $R_B$ in these two paths, where $R_S$ and $R_B$ jointly dominate the integrity loss of a cell. However,  $R_B$ is controlled by the neighboring \textit{access pattern}, therefore, by observing or measuring the $R_B$ we can infer \textit{access pattern} inversely to violate confidentiality. 
}


\subsubsection{P-well Leakage ($I_1$)}
\label{sec:Effective Intrinsic Leakage}



As discussed in Section~\ref{sec:Sensing Noise}, repeated wordline activations
inject minority carriers into the cell-array p-well substrate. The transient
voltage on the aggressor wordline locally increases the free-electron density
in the p-well, and these carriers are subsequently captured by neighboring
cells. As a result, the intrinsic leakage path $I_1$ becomes more conductive,
which we capture by an effective resistance $R_S$.

In contrast to pure retention, the relevant stress parameter is not elapsed time but the cumulative activation history of neighbouring rows. Each activation keeps the wordline high for an effective stress window $t_{\text{AggON}}$ during which additional electrons are injected into the p-well. After
\zl{hammering counts} $(HC)$ such activations, the victim cell has effectively experienced a total
leakage exposure of $t_{\text{eff}} = HC \cdot t_{\text{AggON}}$.

Starting from the standard $RC$ decay for the storage node,
$V_S(t) = V_{DD} \exp(-t/(R_S C_S))$, we reparameterize the evolution in terms
of the hammer count $HC$ by substituting $t = t_{\text{eff}}$. This yields a
count-domain leakage model
\begin{equation}
  I_1(HC)
  = \frac{V_S(HC)}{R_S}
  = \frac{V_{DD}}{R_S} \cdot
    \exp\!\left(
      -\,\frac{HC \cdot t_{\text{AggON}}}{R_S C_S}
    \right).
  \label{eq:I1_model_count}
\end{equation}

\subsubsection{Bitline Leakage ($I_2$)}

\zilong{Bitline leakage $I_2$ indicates the subthreshold current between the source and drain of the access transistor when the channel is not totally opened.}
During an \texttt{ACT} command process, the aggressor wordline rises from $0$ to $V_{\mathrm{PP}}$ within $T_{RCD}$, producing a transient spike on the victim wordline:
\begin{equation}
\Delta V_{\mathrm{victim}} \approx V_{\mathrm{PP}} \cdot \frac{C_C}{C_C + C_{\mathrm{WL}}},
\end{equation}
where $C_C$ is the parasitic capacitance $C_C$ couples adjacent wordlines and $C_{\mathrm{WL}}$ is the victim wordline capacitance to ground.

Although $\Delta V_{\mathrm{victim}}$ does not exceed the threshold $V_T$, it biases the transistor into the subthreshold region, yielding a leakage current $I_2$
\begin{equation}
I_{2}(t) = \mu_n C_d v_{\text{th}}^2 \cdot \frac{W}{L} \cdot 
e^{\frac{\Delta V_{\mathrm{victim}} - V_B - V_T}{n v_{\text{th}}}} 
\left( 1 - e^{-\frac{V_{S(t)} - V_B}{v_{\text{th}}}} \right),
\end{equation}
where $\mu_n$ is electron mobility, $C_d$ the depletion capacitance, $W/L$ the aspect ratio, $v_{\text{th}}$ the thermal voltage, $n$ the subthreshold swing coefficient, $V_S$ the victim storage-node voltage, and $V_B$ the bitline voltage. 
\zilong{To further express effective resistance $R_B$:}
\begin{equation}
\boxed{
R_B(V_S, V_B) =
\frac{V_S - V_B}{
A \left( 1 - e^{-\frac{V_S - V_B}{v_{\mathrm{th}}}} \right)
}.
}
\end{equation}
with the \textbf{Subthreshold Factor} $A$
\begin{equation}
A = \mu_n C_d v_{\text{th}}^2 \cdot \frac{W}{L} \cdot 
e^{\frac{\Delta V_{\mathrm{victim}}- V_B - V_T}{n v_{\text{th}}}},
\end{equation}

\zilong{As shown in Figure \ref{fig:DRAM_cell_2}, $R_B$ through the $I_2$ is controlled by the \emph{access data pattern}. We will discuss the different patterns and how we model predicts as follows:} 

\paragraph{111 pattern case:}
If capacitors in the victim rows and aggressor rows store the same value (e.g., \texttt{111}), during the read disturbance, $V_B$ will be pulled up to $V_{DD}$ by aggressor cells. Therefore, when $V_S \approx V_B$, it makes $\lvert V_S - V_B \rvert \ll v_{\mathrm{th}}$,
the exponential term can be linearized as
$(1 - e^{-x}) \approx x$, leading to
\begin{equation}
R_B^{111} \approx \frac{v_{\mathrm{th}}}{A \cdot (V_S-V_B)}
\end{equation}
\zilong{Consequently, the effective resistance $R_B$ 
increases to \textbf{a very large value}, indicating that the leakage toward the bitline is practically \emph{insignificant} in this case.}
\paragraph{010 pattern case:}
\zilong{In the contrast, when the victim rows and aggressor rows store the opposite vale (e.g., \texttt{010}). $V_B$ will be pulled down to $0$ by aggressor rows under the read disturbance. In this case, $\lvert V_S - V_B \rvert \approx V_{DD} > v_{\mathrm{th}}$ , makes $(1 - e^{-x}) \approx 1$, leading to}

\begin{equation}
R_B^{010} \approx \frac{v_{\mathrm{th}}}{A}
\end{equation}

\zilong{Through this modelling, we find that under the $010$ pattern effective resistance $R_B$ becomes a reasonable value which depends on the \textbf{subthreshold voltage} $v_{\mathrm{th}}$ of access transistor and our defined \textbf{subthreshold factor} $A$. It also reveals the mechanisms of the widely observed phenomenon - why CheckerBoard pattern \texttt{010} is able to accelerate the charge loss of the cell then increase the likelihood of the bitflips. }

\subsection{Case Studies}
\label{subsection:case_studies}
\zilong{Before detailed attack qualification, we present two representative scenarios to illustrate how the proposed $(R_S,R_B)$-based leakage model supports \emph{decision-oriented} reasoning about integrity and confidentiality vulnerabilities.}
They include two attacks aligned with our model-driven adversarial objectives. \textbf{Case~A} targets \textit{integrity} via pattern-sculpted disturbance to edit a \emph{targeted} sensitive bit; \textbf{Case~B} targets \textit{confidentiality} by inferring an unknown neighbour pattern from the flip statistics.

\subsubsection{Case A: Pattern-Sculpted Positioned Edit (Integrity)}
\zilong{This instantiates the precise integrity violation scenario (\textbf{OBJ-I-2}, Pattern Control Attacker) from Section \ref{sec3:threat_model}. The goal is to flip a chosen bit in the DRAM arrays and leave other bits unchanged by increasing its flip likelihood through controlled neighbour access patterns.
In the equivalent model, pattern control primarily acts through the bitline-side path: it reduces $R_B$ under $010$ and increases $R_B$ under $111$.
Let the effective time constant under pattern \texttt{pat} be }
\begin{equation}
\tau^{\texttt{pat}} \;=\; C_S \bigl(R_S \parallel R_B^{\texttt{pat}}\bigr),
\label{eq:tau_pat}
\end{equation}
\zilong{so that, for a fixed stress window (e.g., $t_{\mathrm{AggON}}$ and activation count), the flip probability increases as $\tau$ decreases.}

\paragraph{Selective flipping via $R_S$ and $R_B$.}
\zilong{For a target cell that flips only under 010, pattern control makes $R_B^{010}\!\ll\!R_B^{111}$.
A sufficient condition for selectivity is the existence of a threshold $\tau^\star$ (determined by the stress schedule) such that}

\begin{equation}
\tau_{\mathrm{target}}^{010} = C_S\!\left(R_S^{\mathrm{tar}} \parallel R_B^{010}\right),
\qquad
\tau_{\mathrm{byst}}^{111} = C_S\!\left(R_S^{\mathrm{byst}} \parallel R_B^{111}\right).
\end{equation}

\begin{center}
\fbox{\parbox{0.92\linewidth}{
\centering
\textbf{Selective-flip succeeds if there exists a stress level } $\tau^\star$
\textbf{ such that } 
\[
\boxed{\;\tau_{\mathrm{target}}^{010} \;<\; \tau^\star \;<\; \tau_{\mathrm{byst}}^{111}\;}
\]
(e.g., choose $t_{\mathrm{AggON}}$/$\#$activations that corresponds to $\tau^\star$).
}}
\end{center}

\zilong{When $R_B^{010}\!\ll\!R_S^{\mathrm{tar}}$, the target’s time constant is dominated by the bitline path ($\tau_{\mathrm{target}}^{010}\!\approx\!C_S R_B^{010}$), enabling a high flip likelihood.
Conversely, suppressing bystanders relies not only on a large $R_B^{111}$ but also on sufficiently large $R_S^{\mathrm{byst}}$; otherwise,}
\begin{equation}
R_S^{\mathrm{byst}} \ll R_B^{111}
\;\Rightarrow\;
\tau_{\mathrm{byst}}^{111} \approx C_S R_S^{\mathrm{byst}} \quad\text{(small)},
\end{equation}
and unwanted flips may occur even under 111.

\paragraph{Practical implication.}
\zilong{Based on this analysis in the practical \emph{positioned integrity} attack design, in order to increase the attack success rate, attackers should refer to the following metrics:}

\begin{center}
\resizebox{\linewidth}{!}{
\begin{tabular}{lcc}
\toprule
 & \textbf{Parameter push} & \textbf{Effect} \\
\midrule
\textbf{Target (010)} 
  & $R_B^{010}\!\downarrow$ \quad(\emph{stronger BL path}) 
  & $\tau_{\mathrm{target}}^{010}\!\downarrow$ \\
\midrule
\textbf{Bystanders (111)} 
 & $R_S^{\mathrm{byst}}\!\uparrow$ \quad(\emph{less self-leak}) 
  & $\tau_{\mathrm{byst}}^{111}\!\uparrow$ \\
\bottomrule
\end{tabular}
}
\end{center}
\zilong{Therefore, for practical attackers, their goal is to enlarge the gap $\Delta\tau=\tau_{\mathrm{byst}}^{111}-\tau_{\mathrm{target}}^{010}$ and the gap $\Delta R =R_{S}^{byst}-R_{B}^{010}$ to effectively launch the attacks. }

\subsubsection{Case B: Pattern-to-Flip Tomography (Confidentiality)}
\zilong{This instantiates the precise confidentiality violation scenario (\textbf{OBJ-C-2}, Pattern Inference Attacker) from Section \ref{sec3:threat_model}.}
The goal of the adversary is to infer the unknown data bit of a \emph{neighbour} (victim) cell by observing the flip statistics of a \emph{probe} cell whose neighbourhood includes that victim.
\zilong{We assume the \emph{aggressor row} encodes the unknown bit $d\in\{0,1\}$ to be inferred.
For each victim cell (whose $R_S$ and $R_B^{\texttt{pat}}$ are known), we drive the neighbourhood with two patterns,
$\texttt{pat}\in\{010,111\}$, and read flips on the victim as a probe of $d$.
The victim's pattern-conditioned time constants are defined as}
\begin{equation}
\tau^{\texttt{pat}}
= C_S \bigl( R_S \parallel R_B^{\texttt{pat}} \bigr)
\end{equation}
for a fixed stress window (e.g., $t_{\mathrm{AggON}}$, activation count).
\zilong{The following box indicates the necessary conditions so that the cell flips under 010 and remains stable under 111.}

\begin{center}
\fbox{\parbox{0.92\linewidth}{
\centering
\textbf{Unambiguous inference with one victim cell requires a separable window}
\[
\boxed{\ \tau^{010} \;<\; \tau^\star \;<\; \tau^{111}\ }
\]
When this holds, the decision rule
\(
\text{flip}\Rightarrow \hat d=1,\ \text{no flip}\Rightarrow \hat d=0
\)
is correct for that cell.
}}
\end{center}

\paragraph{Practical implication.}

\zilong{If there is not enough gap between the value of the $R_B^{010}$ and $R_S$, the discharge time-constants under 010 and 111 become comparable, consequently the decision margin (likelihood ratio) collapses and the posterior $\Pr(\text{aggressor}=0\,|\,\text{obs})$ cannot be separated from $\Pr(\text{aggressor}=1\,|\,\text{obs})$ (i.e., stays close to $0.5$).}

\zilong{Therefore, in a practical \emph{confidentiality} attack, in order to develop the accuracy of the inference, attackers should refer to the following metrics:}
\begin{center}
\resizebox{\linewidth}{!}{
\begin{tabular}{lcc}
\toprule
\textbf{Parameter push} & \textbf{Effect on $\tau^{010}$, $\tau^{111}$} & \textbf{Net impact} \\
\midrule
$R_B^{010}\!\downarrow$ & $\tau^{010}\!\downarrow$ & easier 010 flip \\
$R_S\!\uparrow$          & both $\tau^{010},\tau^{111}\!\uparrow$ & harder 111 flip \\
\bottomrule
\end{tabular}
}
\end{center}

\zilong{Similarly, the goal of the \emph{confidentiality} attackers is to enlarge the gap $\Delta\tau=\tau_{\mathrm{probe}}^{111}-\tau_{\mathrm{probe}}^{010}$ and the gap $\Delta R =R_{S}-R_{B}^{010}$ to effectively launch the attacks. }

\zilong{Taken together, Case~A and Case~B show how our $(R_S,R_B)$-based leakage model directly links device-level parameters to the success conditions of both targeted integrity edits and pattern-inference attacks. By reading off gaps such as $\Delta\tau$ and $\Delta R$ from our calibrated parameters, an adversary can systematically select the most exploitable cells and stress windows. We will give a more detailed qualification of attacks in the following section.}

\subsection{Attacks Qualification}
\label{sec:Attacks Modeling}
\zilong{After two case studies, we adopted the parameters in our model to precisely explain the integrity and confidentiality attacks quality in three classes of attack scenarios considered in the Threat Model (Section \ref{sec3:threat_model}):  }
\paragraph{\textbf{(i) General integrity attacks} (\textbf{OBJ-I-1, OBJ-I-2, OBJ-V-1, OBJ-V-2}).}
The attacker’s goal is to \emph{flip stored bits without direct writes}, i.e., to drive the storage node across the sense-amplifier decision threshold during a read. In our model, the vulnerability of cells is governed by the effective leakage resistance $R_S$ and $R_B$. For a given mechanism and pattern, the total leakage conductance can be written as

\begin{equation}
 G_{\text{tot}} = \frac{1}{R_S} + \frac{1}{R_B}   
\end{equation}
  
where $R_S$ captures the cell-to-well leakage and $R_B$ aggregates any additional bitline-side leakage branch (if present). The corresponding time constant is $\tau \approx C_S / G_{\text{tot}}$, so a larger $G_{\text{tot}}$ directly translates into shorter flip times and stronger integrity attacks.

Pure retention failures operate with $R_B \to \infty$, such that $G_{\text{tot}} \approx 1/R_S$. Rowhammer and Rowpress both add a finite bitline-side branch $R_B$ and hence increase $G_{\text{tot}}$ under the access pattern $010$.
Therefore, the quality of a general integrity attack can be determined by the total leakage conductance $G_{\text{tot}}$.

\paragraph{\textbf{(ii) Targeted integrity attacks} (\textbf{OBJ-I-2})}
Beyond flipping bits as fast as possible, a more powerful adversary aims for \emph{selective} integrity violations: inducing flips only under specific data patterns while keeping other patterns stable. In this case, the selectivity is not determined by the absolute leakage $G_{\text{tot}}$ alone, but by the \emph{pattern-induced difference} between leakage under two patterns, e.g.,
010 vs.\ 111. For a victim cell $i$, we write
\begin{equation}
\label{equation: G111}
  G_{111}(i) \approx \frac{1}{R_S(i)}, \quad
  G_{010}(i) = \frac{1}{R_S(i)} + \frac{1}{R_B(i)} ,
\end{equation}
so that the additional pattern lever is
\begin{equation}
\label{equation:Delta G(i)}
  \Delta G(i) \triangleq G_{010}(i) - G_{111}(i) = \frac{1}{R_B(i)}
\end{equation}

A small $R_B$ (large $1/R_B$) means that the 010 pattern decays faster than 111. However, true selectivity also requires that the baseline leakage under 111 remains small; otherwise both patterns may flip within the same stress window.

To capture this, we consider the \emph{relative} pattern gap
\begin{equation}
\label{eq:relative_G}
     \Delta G_{\mathrm{rel}}(i)
  \triangleq \frac{G_{010}(i) - G_{111}(i)}{G_{111}(i)}
  = \frac{1/R_B(i)}{1/R_S(i)}
  = \frac{R_S(i)}{R_B(i)} 
\end{equation}

In this view, a large $R_S(i)$ suppresses leakage under 111 while a small
$R_B(i)$ strongly enhances leakage under 010; both are jointly necessary for highly selective integrity attacks. 

\paragraph{\textbf{(iii) Confidentiality attacks. }(\textbf{OBJ-C-2})}
In confidentiality attacks, the adversary does not aim to corrupt data but to
\emph{infer} a hidden pattern (e.g., whether the neighborhood stores 0 or 1)
from the flip statistics of a victim row. The attacker typically adopts a
simple rule: if a flip is observed within the attack budget, guess the
``vulnerable'' pattern (e.g., 010, storing 0); if no flip is observed, guess
the ``stable'' pattern (e.g., 111, storing 1).

\zilong{Similar to the targeted integrity attack, it is the \emph{gap} between the intrinsic p-well resistance $R_S$ and the pattern-dependent bitline resistance $R_B$ that drives confidentiality. For each bit $i$, the effective conductances under 111 and 010 patterns are already defined in Equation~\eqref{equation: G111}.}
\zilong{For a fixed hammer budget, we abstract the stress as a conductance threshold $\theta$: bits with $G_{010}(i) \ge \theta$ will flip with high probability under the 010 pattern, and bits with $G_{111}(i) \ge \theta$ will flip with high probability under the 111 pattern. Let $p_{010}(\theta)$ and $p_{111}(\theta)$ denote the fractions of bits that flip under 010 and 111, respectively. Under the rule “flip $\Rightarrow 0$, no flip $\Rightarrow 1$” and assuming equal priors, the expected inference accuracy at stress level $\theta$ is}
\begin{equation}
  \widehat{\mathrm{Acc}}(\theta)
  = \frac{1}{2}\Bigl[p_{010}(\theta) + \bigl(1 - p_{111}(\theta)\Bigr)\Bigr].
  \label{eq:acc_theta}
\end{equation}
\section{Experiments and Attack Evaluations}
\label{sec5:evaluation}
\vspace{-1.2mm}
\subsection{Experimental Objectives and Overview}

\zilong{The purpose of this section is to empirically validate that the cell-level leakage model in
Section~\ref{sec4:memory_model} provides \emph{actionable, decision-oriented} insights for the
security scenarios formalized in Section~\ref{sec3:threat_model}. In particular, this section
addresses the three research questions (RQs) introduced in
Section~\ref{sec:motivation} from an empirical perspective.}


\textbf{RQ1 (Measurability of physical parameters).}  
We first validate that the proposed circuit-level parameters—effective p-well resistance ($R_{S}$), effective bitline resistance ($R_{B}$), pattern-induced noise ($N$) can be reliably extracted from real DRAM modules. Through systematic retention profiling and read disturbance (Rowhammer and Rowperss) stress tests across multiple devices, we examine whether these quantities are measurable in practice.
This confirms that our framework is not only theoretical but also experimentally grounded.  

\textbf{RQ2 (Security Analysis of attack surfaces).}  
\zilong{Given the extracted parameters, we then study how they can be used to analyse and qualify the performance of integrity and confidentiality attacks. We also compare the parameters across different DRAM chips as well as different read disturbance mechanisms (Eg, Rowhammer and Rowpress).}

\textbf{RQ3 (Insights on existing attack surfaces).}  
\zilong{Finally, we investigate whether Rowpress can strengthen or extend existing attack surfaces when compared to Rowhammer under a comparable ``damage budget'' (e.g., total flip count or stress window). We focus on two aspects: (i) selective integrity violation, where the attacker aims to steer flips towards specific victim cells while minimizing collateral damage, and (ii) pattern-dependent confidentiality leakage, where the attacker infers unknown data through disturbance-induced bitflips. 
Our experiments show that Rowpress exposes finer-grained, more pattern-aware control knobs, suggesting an expanded security risk surface that is not captured by prior empirical-only analyses.}

\vspace{-1.2mm}
\subsection{Hardware Platform}
\begin{algorithm}[t]
\caption{Extraction of $R_S^{\mathrm{RD}}$ and $R_B^{\mathrm{RD}}$ from read disturbance (Rowpress and Rowhammer)}
\label{alg:reffA}
\begin{algorithmic}[1]
\small 
\State \textbf{Input:} cell $v$; patterns $P\in\{\texttt{111},\texttt{010}\}$;
hammer rate $f_{\mathrm{RD}}$; target flips $N_{\mathrm{target}}$;
$C_S$, $V_{DD}$, $V_{\mathrm{flip}}$, $V_{TH}$
\State \textbf{Output:} $R_S^{\mathrm{RD}}$, $R_B^{\mathrm{RD}}$

\Function{HCAtTargetFlips}{$v$, $P$}
  \State $HC \gets HC_{\mathrm{init}}$
  \While{victim flips $< N_{\mathrm{target}}$}
    \State Configure victim/aggressors for pattern $P$
    \State Initialize $v \gets \texttt{1}$
    \State Hammer aggressors with count $HC$ at rate $f_{\mathrm{RD}}$
    \State Read victim and count flips
    \If{flips $< N_{\mathrm{target}}$} \State $HC \gets \alpha \cdot HC$ \Comment{e.g., $\alpha=2$} \EndIf
  \EndWhile
  \State \Return $HC$
\EndFunction

\State $HC_{111} \gets$ \Call{HCAtTargetFlips}{$v$, \texttt{111}}
\State $HC_{010} \gets$ \Call{HCAtTargetFlips}{$v$, \texttt{010}}
\State $T_{111} \gets HC_{111}/f_{\mathrm{RD}}$, \quad $T_{010} \gets HC_{010}/f_{\mathrm{RD}}$

\State $R_{\mathrm{eff}} \gets \dfrac{T_{111}}{C_S \ln\!\big(\tfrac{V_{DD}}{V_{\mathrm{flip}}}\big)}$ \Comment{$R_{\mathrm{eff}}\approx R_S^{\mathrm{RD}}$}
\State $E \gets \big(\tfrac{V_{DD}}{V_{\mathrm{flip}}}\big)^{T_{010}/T_{111}}$
\State $A \gets \dfrac{1}{R_{\mathrm{eff}}}\cdot\dfrac{V_{DD}-E V_{\mathrm{flip}}}{E-1}$
\State $R_S^{\mathrm{RD}} \gets R_{\mathrm{eff}}$, \quad $R_B^{\mathrm{RD}} \gets \dfrac{V_{TH}}{A}$
\State \Return $R_S^{\mathrm{RD}}, R_B^{\mathrm{RD}}$
\end{algorithmic}
\end{algorithm}
In this section, we present the hardware \zilong{platform} of our test framework, including the FPGA-based experimental platform, the custom memory controller, and the diverse set of DRAM chips used for validation.  



\textbf{Experimental platform:}
We conduct our experiments on an Xilinx Alveo U200 FPGA accelerator card, which can work with high-bandwidth DDR4 DRAM modules
. 
A custom memory controller is implemented on the FPGA, enabling fine-grained control over the memory under test.

\textbf{Memory Controller:}
For this, we leverage \textit{DRAM Bender}\cite{olgun2023dram}, a flexible open-source framework that enables direct manipulation of DRAM timing parameters and access patterns. DRAM Bender allows us to bypass conventional memory controllers and issue raw DRAM commands (e.g., \texttt{ACT}, \texttt{PRE}, \texttt{RD}, \texttt{WR}) directly to the device. This low-level programmability makes it possible to construct customized access sequences, such as hammering patterns or controlled retention tests, and to precisely measure resulting bitflips. \zl{In addition, the customized memory controller follows timing configurations under the JEDEC DDR4 standard \cite{jedec_ddr4_2020}.
Specifically, $t_{RCD}$ and $t_{RP}$ were set to 13.5\,$ns$, while $t_{RAS}$ was set to 35\,$ns$ (except for Rowpress test).}

\textbf{DRAM Chips:}
To validate our cell-level circuit models, we selected a diverse set of DRAM modules that span multiple sources of variability.  
\zl{The detailed information of tested DRAM modules is shown in the Table \ref{tab:dimms}.}
This set includes (i) multiple modules of the same model from a single vendor (Micron 16\,GB) to capture device-to-device variation under nominally identical fabrication;  
(ii) modules with different density and scaling generations (Micron 16\,GB, Lenovo 8\,GB vs ADATA 4\,GB) to reflect broader technology trends.
(iii) modules from different vendors (ADATA 4\,GB, Innodisk 4\,GB and Lenovo 8\,GB) to compare vendor-specific circuit and process designs.  
Together, these modules provide a comprehensive and representative basis for validating the robustness and generality of our proposed models. It should be noted that our model focuses on the fundamental $1T1C$ cell architecture and its intrinsic leakage mechanisms (p-well and bitline leakage), which remain structurally consistent from DDR4 to DDR5. Of these two DDR4 is currently more practical for these kind of deep-dive empirical tests requiring custom memory controllers and varied test hardware. Thus, the experiments in this paper are based on DDR4 chips.

\begin{table}[t]
\centering
\caption{\zl{Tested DRAM modules}}
\label{tab:dimms}
\resizebox{\columnwidth}{!}{
\begin{tabular}{lcccccc}
\toprule
DIMM & Vendor & Type & Density & Organization & Speed & Year \\
\midrule
\textbf{D1} & Micron & DDR4 & 16GB & 1Rx4 & 2400 & 2021 \\
\textbf{D2} & Micron & DDR4 & 16GB & 1Rx4 & 2400 & 2021 \\
\textbf{D3} & Micron & DDR4 & 16GB & 1Rx4 & 2400 & 2021 \\
\textbf{D4} & Lenovo & DDR4 & 8GB & 1Rx4 & 2666 & 2018 \\
\textbf{D5} & Lenovo & DDR4 & 8GB & 1Rx4 & 2666 & 2018 \\
\textbf{D6} & Innodisk & DDR4 & 8GB & 1Rx4 & 2400 & 2019 \\
\textbf{D7} & ADATA & DDR4 & 4GB & 1Rx4 & 2400 & 2018 \\
\bottomrule
\end{tabular}
}
\end{table}

\subsection{Experimental Results}

We conducted experiments on commodity DDR4 DRAM modules using our FPGA-based testing platform.
For \textit{retention profiling}, we extended the refresh-pausing window up to \textbf{3600 seconds}, for \textit{Rowhammer}, we stressed the cells with double-sided maximum to \textbf{15M hammering counts (HC)}, \zilong{and for \textit{double-sided Rowpress}, we set the $t_{AggON}$ to $1000ns$, then stressed the cells with maximum to \textbf{1.5M hammering counts (HC)}}. Both tests were with different written patterns (baseline \texttt{all-1} and \textit{checkerBoard}) or aggressor-victim patterns (baseline \texttt{111} and \texttt{010}) to capture pattern-dependent effects \zilong{ on the true cells}.
During these experiments, we collected both the total bitflip counts and the spatial distribution maps. By applying Algorithm~\ref{alg:reffA} and Algorithm~\ref{alg:measure_rs_n} (\textbf{presented in Appendix~C}), we extracted the p-well effective resistance $R_{S}$, $R_{S}(N)$(\zl{with} noise) and the bitline effective resistance $R_{B}$ corresponding to \textit{retention failures} and \textit{read} disturbance conditions.
\begin{table*}[ht]
\centering
\resizebox{\textwidth}{!}{%
\begin{threeparttable}
\caption{Measured resistance ranges for DRAM modules}
\label{tab:dram_modules}
\begin{tabular}{lcccccc}
\toprule
\multirow{2}{*}{\textbf{DIMM}} &
\multicolumn{2}{c}{\textbf{Volatility}} &
\multicolumn{2}{c}{\textbf{Rowhammer Test}} &
\multicolumn{2}{c}{\textbf{Rowpress Test}} \\
\cmidrule(lr){2-3} \cmidrule(lr){4-5} \cmidrule(lr){6-7}
& $R_S$ ($\times 10^{13}\,\Omega$) 
& $R_S(N)$ ($\times 10^{13}\,\Omega$)
& $R_S$ ($\times 10^{10}\,\Omega$) 
& $R_B$ ($\times 10^{8}\,\Omega$)
& $R_S$ ($\times 10^{10}\,\Omega$) 
& $R_B$ ($\times 10^{8}\,\Omega$) \\
\midrule
\textbf{D1} & 10.6--5530 & 9.01--5244 & 4.78--305   & 57.9--909  & 36.1--1050 & 38.5--362 \\
\textbf{D2} & 9.2--4988  & 7.7--4620  & 5.31--306   & 37.2--905  & 25.5--2125 & 25.7--584 \\
\textbf{D3} & 13.2--6720 & 11.4--5420 & 5.31--240   & 76.7--829  & 55.3--1830 & 44.4--258 \\
\textbf{D4} & 8.50--8960 & 7.2--6980  & 4.54--31880 & 29.5--6470 & 102--638000 & 24.3--1170 \\
\textbf{D5} & 6.5--7350  & 5.5--6321  & 5.21--46300 & 36.2--8410 & 512--826000 & 15.6--2230 \\
\textbf{D6} & 14.4--5620 & 10.8--4770 & 8.77--442   & 44.3--486  & 25.5--1620 & 9.45--312 \\
\textbf{D7} & 15.6--6820 & 13.2--5183 & 4.31--85.0  & 37.8--643  & 31.9--424 & 26.7--393 \\

\bottomrule
\end{tabular}
\end{threeparttable}
}
\end{table*}

\vspace{-1.2mm}
\subsubsection{\emph{w.r.t.} RQ1: Measurability of Physical Parameters}
 
We first verify whether the proposed parameters $R_{S}$, $R_{B}$, and $N$ can be reliably extracted from real DRAM modules, and whether their statistical properties are consistent with our models.
\zilong{Table~\ref{tab:dram_modules} summarizes the tested DRAM modules and the extracted parameters of $5000$ most vulnerable \textit{true cells} in each row of each memory chip in our study.  For each device, we report: 
(i) the p-well effective resistance $R_{S}$ estimated under the all-\texttt{1} retention pattern,  
(ii) the effective noise parameter $R_{S}(N)$ obtained from the checkerboard retention pattern,  
(iii) the p-well effective resistance $R_{S}$ and bitline effective resistance $R_{B}$ measured under Rowhammer  and Rowpress with the \texttt{111} and \texttt{010} activation pattern.}

To enable security reasoning in later sections, we summarise two empirically robust observations that directly reflect the physical structure of DRAM cell leakage and determine which attack surfaces are practically viable.

\vspace{-\parskip}


\textcolor{black}{\textbf{Observation 1.}}
\zilong{\textit{Rowpress-induced p-well effective resistance $R_{S}^{RP}$ is consistently orders of magnitude greater than the Rowhammer-induced p-well effective resistance $R_{S}^{RH}$, revealing Rowpress induces much smaller leakage to the p-well than Rowhammer.}}
\vspace{-\parskip}

\zilong{We observe that for every tested  Dual In-Line Memory Module (DIMM), the Rowpress p-well resistance range $R_S^{\mathrm{RP}}$ is shifted to significantly larger values than the corresponding Rowhammer range $R_S^{\mathrm{RH}}$. For instance, on Micron$_1$ the Rowhammer $R_S^{\mathrm{RH}}$ spans $4.78$–$305 \times 10^{10}\,\Omega$, whereas Rowpress reaches $36.1$–$1050 \times 10^{10}\,\Omega$; similarly, Lenovo$_1$, Lenovo$_2$, Innodisk, ADATA and Team all show $R_S^{\mathrm{RP}}$ that are roughly one to two orders of magnitude higher than $R_S^{\mathrm{RH}}$, confirming that Rowpress induces much weaker effective p-well leakage than Rowhammer in practice. This implies that, in the absence of any pattern-controlled bitline effect, Rowpress does aggravate the intrinsic cell vulnerability; however, it is in fact \emph{less} damaging to the p-well leakage path than Rowhammer.}

\vspace{-\parskip}
\textcolor{black}{\textbf{Observation 2.}}
\zilong{\textit{Rowpress-induced bitline effective resistance $R_{B}^{RP}$ is consistently orders of magnitude smaller than the Rowhammer-induced bitline effective resistance $R_{B}^{RH}$, revealing Rowpress induces much stronger leakage to the bitline than Rowhammer.
}}
\vspace{-\parskip}

\zilong{From Table~\ref{tab:dram_modules}, we see that for all tested DIMMs, the Rowpress $R_{B}^{RP}$ ranges (in $10^{8}\,\Omega$) are systematically shifted to much lower values than the corresponding Rowhammer $R_{B}^{RH}$ ranges, often by one to two orders of magnitude. This implies that, once a suitable 010-like access pattern is applied, Rowpress can dominantly route the discharge through the bitline branch and thus offer attackers a substantially stronger handle for both targeted integrity edits and pattern-dependent confidentiality attacks than Rowhammer.}

\subsubsection{\emph{w.r.t.} RQ2: Measured Parameters for Attack Qualification}
\label{sec:w.r.t RQ2}
\begin{figure}[t]
  \centering
  \begin{subfigure}{\linewidth}
    \centering
    \includegraphics[width=0.9\linewidth]{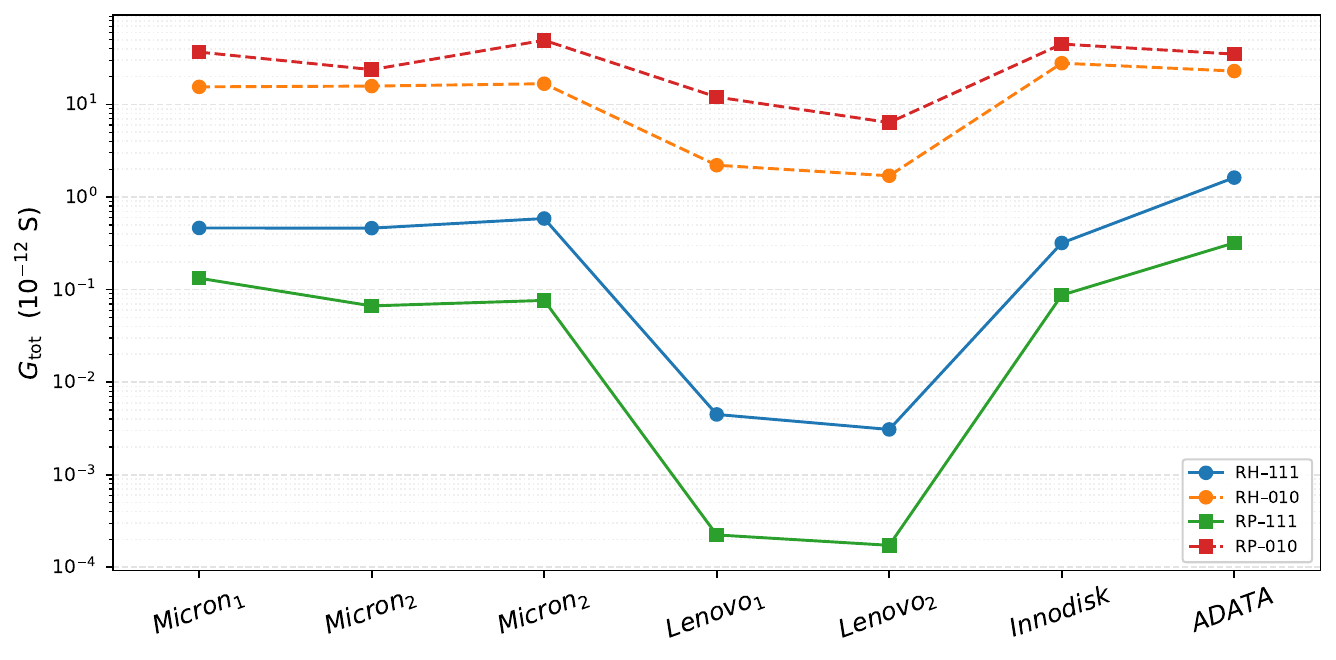}
  \end{subfigure}
  \caption{The total leakage conductance $G_{\text{tot}}$ under Rowhammer vs Rowpress.}
  \label{fig:G_tot}
\end{figure}

\zilong{So far, we have extracted the leakage parameters $(R_S, R_B)$ from both retention-failure and read-disturbance experiments and reported several observations on how different mechanisms and patterns shape the attack surface. We now answer RQ2 by adopt these parameters to directly qualify the attacks by the metrics introduced in Section \ref{sec:Attacks Modeling}.}
\vspace{-\parskip}
\paragraph{(i) General integrity attacks.}
\zilong{Our \textbf{Observation 1} and \textbf{Observation 2} already implied that Rowhammer has stronger leakage than Rowpress through $R_S$ and weaker leakage through $R_B$. Intuitively, it indicates that Rowhammer performs a stronger intrinsic leakage than Rowpress, but Rowpress induces more pattern-controlled leakage.  
For a more detailed quantification of their impacts, we can take use of the Equation \ref{fig:G_tot} to calculate and compare.}

For each victim cell $i$, we quantify the value of leakage conductance to evaluate direct integrity attacks under Rowpress over Rowhammer in different access patterns.
\zilong{In Figure~\ref{fig:G_tot},  we aggregate $G_{\text{tot}}(i)$ over all test rows to obtain per-module medians across our test DRAM chips. Across modules, the RH–111 curve stays above RP–111, indicating stronger integrity capability when only $R_S$ is active, while under the 010 pattern the RP–010 curve is significantly elevated and \zl{always} surpasses RH–010, showing that once bitline leakage is enabled, Rowpress provides a much stronger direct integrity attack handle than Rowhammer.}

\begin{notebox}
Rowhammer yields a larger attack capability than Rowpress under the 111 pattern, but once the 010 pattern enables a low-$R_B$ bitline path, Rowpress overtakes Rowhammer and becomes the more powerful general integrity \zl{attack vector}.
\end{notebox}

\paragraph{(ii) Targeted integrity attacks}
\zilong{Our two key observations also show, in general, Rowpress has a larger gap between $R_S$ and $R_B$ than Rowhammer. Therefore, it means higher flipping selectivity if attackers tend to control bit flips precisely through access patterns. For further precise measurement of the attack quality, in section \ref{sec:Attacks Modeling}, we provide \textit{relative} pattern lever $\Delta G_{\mathrm{rel}}(i)$ (Equation \ref{eq:relative_G}) to measure the attacker's capability of targeted integrity violation.}

\begin{figure}[t]
  \centering
  \begin{subfigure}{\linewidth}
    \centering
    \includegraphics[width=0.9\linewidth]{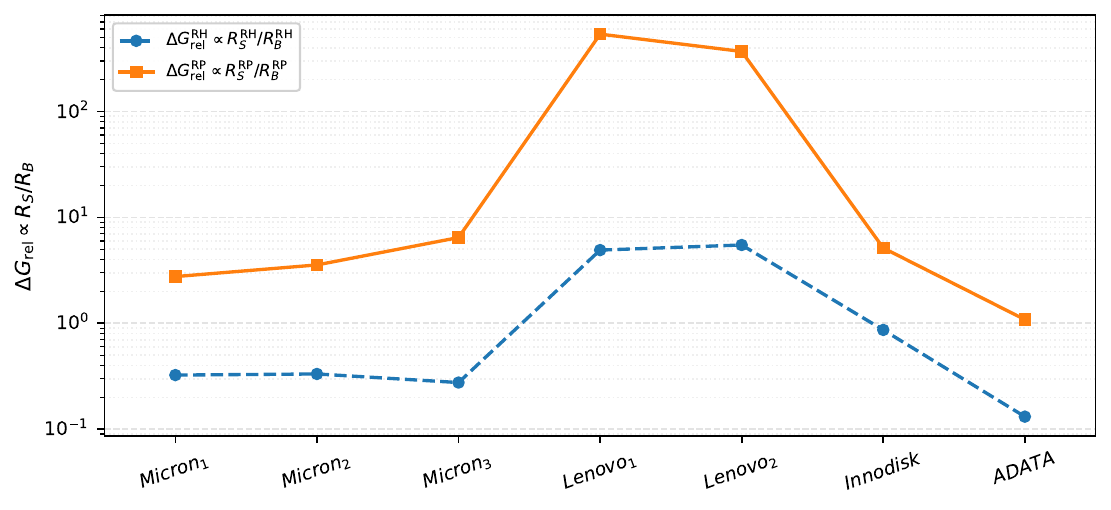}
    \label{fig:DeltaG}
  \end{subfigure}\hfill
  \caption{Relative Pattern Lever $\Delta G_{\mathrm{rel}}$  under Rowhammer vs Rowpress.}
  \label{fig:DeltaG_K_tot}
\end{figure}

\zilong{As illustrated in Figure \ref{fig:DeltaG_K_tot}, we aggregate $\Delta G_{\mathrm{rel}}(i)$ over all test rows to obtain per-module medians across our test DRAM chips.} \zilong{It shows that, for every DIMM, Rowpress systematically yields much larger $\Delta G_{\mathrm{rel}}$ than Rowhammer, often by more than an order of magnitude. It indicates that the pattern lever between vulnerable (010) and stable (111) configurations is significantly stronger under Rowpress and thus more favorable for targeted integrity attacks. }




\begin{notebox}
Across all tested DIMMs, Rowpress consistently provides a much stronger capability for targeted integrity attacks than Rowhammer, enabling more selective and controllable bit flips under the same disturbance budget.
\end{notebox}

\paragraph{(iii) Confidentiality attacks.}
\zilong{To capture the \emph{practical} confidentiality attack capability on real DIMMs, we instantiate this model using the extracted per-row parameters $(R_S, R_B)$ for both Rowhammer and Rowpress. For each mechanism and each DIMM, we rank all bits by their 010-pattern conductance and select the $m_{010}$ most vulnerable bits such that, under a fixed hammer budget $HC$, all of them flip when the adjacent pattern is 010. Applying the same $HC$ when the adjacent pattern is 111, we count how many of these $m_{010}$ bits also flip and denote this number by $m_{111}$.}

\zilong{The attacker keeps the same decision rule: if a bit flips, guess 010; otherwise, guess 111. For the selected $m_{010}$ bits, the success probability when the true pattern is 010 is $1$ (they all flip by construction), while the success probability when the true pattern is 111 is $1 - m_{111}/m_{010}$. Assuming equal priors, the overall estimated inference accuracy becomes}
\begin{equation}
  \mathrm{Acc}
  = \tfrac{1}{2}\Bigl[1 + \bigl(1 - \tfrac{m_{111}}{m_{010}}\bigr)\Bigr]
  = 1 - \tfrac{1}{2}\,\frac{m_{111}}{m_{010}}.
  \label{eq:acc_m010_m111}
\end{equation}



\begin{figure}[t]
  \centering
  \begin{subfigure}{\linewidth}
    \centering
    \includegraphics[width=0.9\linewidth]{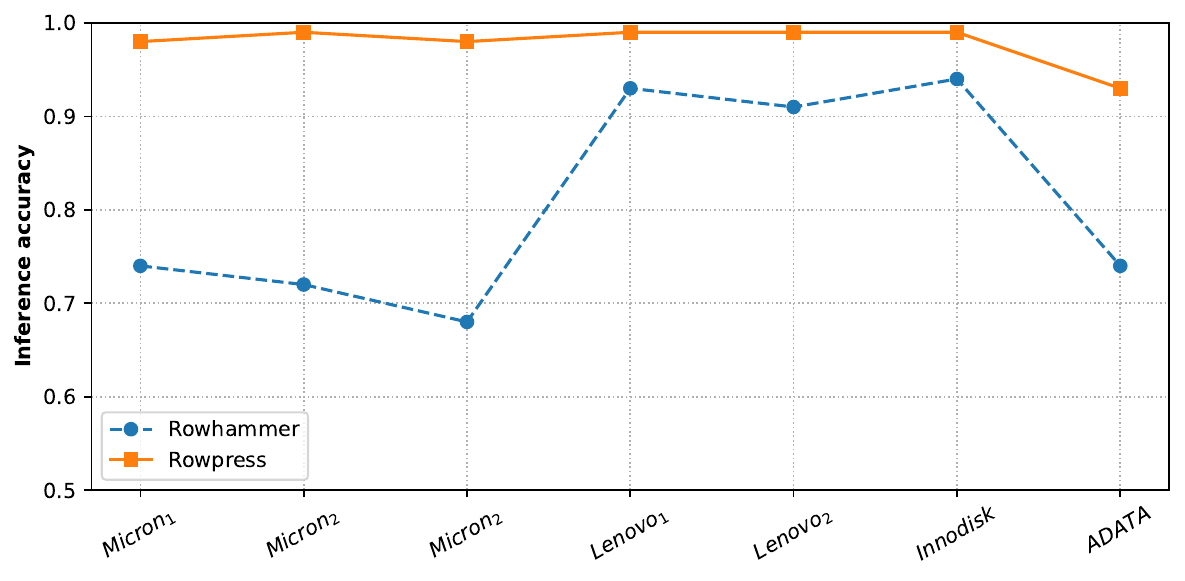}
  \end{subfigure}
  \caption{Estimated average confidentiality inference accuracy among the most vulnerable 5000 cells in each row across DIMMs.}
  \label{fig:acc_dram}
\end{figure}

\zilong{As illustrate in Figure \ref{fig:acc_dram},}
\zilong{we set $m_{010}$ to 5000 and estimate the average inference accuracy by Equation (\ref{eq:acc_m010_m111}) across the test rows and devices. Across all tested DIMMs, Rowpress consistently yields a higher $\mathrm{Acc}_{\max}$ for more aggressive targets (e.g., $>90\%$ accuracy), while Rowhammer often fails to reach the same level within any stress window, highlighting the stronger confidentiality threat enabled by Rowpress.}

\begin{notebox}
Rowpress consistently enables substantially stronger confidentiality attacks than Rowhammer: under the same hammer budget, it achieves higher estimated pattern–inference accuracy and exposes a larger fraction of bits whose values can be reliably inferred from flip behavior.
\end{notebox}

\subsubsection{\emph{w.r.t}. RQ3: Insights on Existing Attacks}
\label{sec:rq3-enhancement}
While RQ2 demonstrates that the extracted $(R_S,R_B)$ parameters accurately characterize integrity and confidentiality vulnerabilities, RQ3 asks a broader system-level question: \emph{Does Rowpress merely replicate Rowhammer, or does it systematically amplify existing Rowhammer-style attacks under identical resource constraints?} Using our calibrated parameters, we compare the two mechanisms across all security-relevant dimensions.

\textbf{First}, for \emph{direct integrity attacks}, we aggregate the total leakage conductance $G_{\text{tot}} = 1/R_S + 1/R_B$ for different mechanisms and patterns (Figure~\ref{fig:G_tot}). Across all DIMMs, the RH–111 curve stays above RP–111, confirming that Rowhammer provides a stronger integrity handle when only the p-well path is active. However, once the 010 pattern enables a low-$R_B$ bitline branch, the RP–010 curve is significantly elevated and often surpasses RH–010, showing that Rowpress becomes the more powerful direct integrity weapon when attackers can exploit pattern-dependent bitline leakage.

\textbf{Second}, for \emph{selective integrity attacks}, we focus on the pattern-induced component $\Delta G = 1/R_B$ and its relative gap $\Delta G_{\mathrm{rel}}(i) = R_S(i)/R_B(i)$. Figure~\ref{fig:DeltaG_K_tot} shows that Rowpress consistently yields much larger $\Delta G_{\mathrm{rel}}$ than Rowhammer across our DIMMs. In other words, Rowpress not only accelerates flips, but also \emph{sharpens} the contrast between vulnerable (010) and stable (111) patterns on the same physical row, making it substantially easier for an adversary to steer bitflips toward targeted data values while keeping bystanders intact.

\textbf{Third}, for \emph{confidentiality attacks}, we instantiate the pattern-inference attacker using the measured $(R_S,R_B)$ and the accuracy model in Equation~(\ref{eq:acc_m010_m111}). For each mechanism and DIMM, we select the $m_{010}$ most vulnerable bits under the 010 pattern, choose a hammer budget $HC$ such that all of them flip under 010, and then measure $m_{111}$—how many of these bits also flip under 111 with the same $HC$. The resulting estimated inference accuracy $Acc = 1 - \tfrac{1}{2} m_{111}/m_{010}$ is summarized in Figure~\ref{fig:acc_dram}. Across all tested DIMMs, Rowpress consistently achieves higher $Acc$ than Rowhammer, and is often the only mechanism that can reach aggressive targets (e.g., $>90\%$ accuracy) within any stress window, highlighting a strictly stronger confidentiality threat.

Overall, the RQ3 analysis shows that Rowpress behaves as a systematic \textbf{amplifier} of existing Rowhammer-style attacks under realistic constraints. It increases $G_{\text{tot}}$ when pattern-controlled bitline leakage is available (direct integrity), enlarges the pattern lever $\Delta G_{\mathrm{rel}}$ (selective integrity), and enables significantly higher inference accuracy $Acc$ for pattern-based key extraction (confidentiality) all without introducing any fundamentally new primitive beyond keeping a row activated for a longer dwell time ($t_{\text{AggON}}$). This suggests that future DRAM defences must explicitly reason about Rowpress-like attacks, rather than focusing solely on traditional Rowhammer activation patterns.


\section{\zl{Impact of In-DRAM Mitigations}}
\label{sec:mitigation}
\zl{Modern DRAM devices may incorporate mitigation mechanisms such as Target Row Refresh (TRR) and on-die Error Correcting Codes (ECC) to reduce the effectiveness of disturbance-based attacks. These mechanisms can significantly affect the practical exploitability of disturbance-induced bit flips.}

\zl{TRR attempts to detect aggressive row activations and refresh nearby victim rows to suppress the accumulation of disturbance effects. As a result, it effectively limits the disturbance budget that an attacker can accumulate within a given time window. However, TRR does not modify the intrinsic electrical properties of the memory cells themselves. Consequently, the resistance parameters extracted in our work ($R_S$ and $R_B$) still characterize the underlying physical vulnerability of the DRAM cells. In practice, TRR reduces the accessible disturbance strength rather than altering the resistance parameters governing bit-flip behaviour.}

\zl{Similarly, on-die ECC is designed to correct a limited number of bit errors before they are exposed to the system. While this mechanism can mask some disturbance-induced errors at the system level, our experimental setup measures raw bit flips directly using an FPGA-based controller. Therefore, the measured resistance parameters reflect the intrinsic device behavior rather than the effectiveness of error correction.}

\zl{Overall, these mitigation mechanisms primarily affect the feasibility of system-level exploitation by constraining disturbance accumulation or masking a subset of errors. However, our model focuses on the characterization of the underlying cell-level vulnerability and provides a quantitative framework for reasoning about disturbance feasibility under different disturbance budgets and mitigation constraints.}

\section{Conclusion}
\label{sec6:conclusion}
This work introduces the \textbf{first} empirically verified cell-circuit-level model that explicitly explains DRAM retention and read–disturbance failures from a security perspective. By modeling the p-well and bitline leakage paths together with pattern-dependent coupling, we extract a small set of physical parameters $(R_S, R_B)$ that directly govern different attack surfaces. Through case studies and attacks qualification, we show how these parameters can be systematically adopted to analyze and construct both \emph{integrity} and \emph{confidentiality} attacks, rather than treating bitflips as opaque symptoms. Implemented and validated on an FPGA-based platform with commodity DRAM modules, our framework provides a quantitative basis for assessing the exploitable surface of real devices to attacks and for comparing mechanisms such as Rowhammer and Rowpress under a unified lens. It offers practical tools to evaluate, prioritize, and ultimately strengthen future DRAM-based designs and defenses.

\bibliographystyle{IEEEtran}
\bibliography{reference}
\appendices
\section{DRAM Organization}

Figure \ref{fig:dram_org} shows an overview of how DRAM is structurally organised. DRAM consists at the lowest level as a set of independent matrices - each known as a bank - of capacitive bit-cells, whereby each bit-cell contains a measurable charge, denoting a logical one or zero as data stored there.  Each bit-cell is a structure known as a 1T1C (one transistor, one capacitor) structure, and consists of five key components: a Cell Capacitor, an Access Transistor, a Storage Node Contact (SNC), a Wordline (WL), a Bitline (BL) and a Bitline Contact (BLC). The cell capacitor stores the actual data bit as electrical charge, with its top plate connected to the access transistor through the SNC, which facilitates the electrical pathway between the two components. The access transistor acts as a switch controlled by the WL signal at its gate terminal, with one source/drain terminal connected to the capacitor via the SNC and the other connected to the BL through the BLC. When the WL is activated during a read or write operation, the access transistor activates, creating a conductive path that allows charge to flow between the capacitor and the BL - either sensing the stored charge (read) or storing a new charge (write). The BLC serves as the connection point between the transistor and the vertical BL that runs through all cells in a column, enabling analogue circuits called sense amplifiers to detect the small voltage changes during reads or to drive new values during writes, which in turn acts as a bridge to transfer data to/from the main memory controller to then be used by CPU. Analogue circuits known as row and column decoders select specific bit-cells to read/write to based on row and column addresses supplied by a main memory controller. Multiple banks (typically 8 to 16) are organised into a single DRAM chip. The smallest form factor seen on any given computer system would be an individual DRAM chip (typically the case with embedded systems), however in commodity/high-powered systems, multiple DRAM chips are found organised across a printed circuit board called a Dual In-Line Memory Module (DIMM). These enable simultaneous chip control across a set of ranks, where each share control signals, yet have separate data pathways. This parallel operation enables the memory controller to interleave read/write accesses across individual chips/banks, allowing DRAM to maintain a very high bandwidth and efficiency, even when individual banks are busy handling operations across multiple clock cycles. 

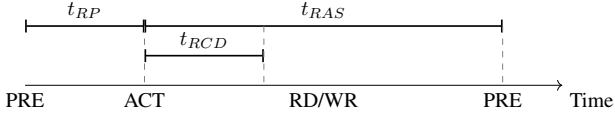
\begin{figure}[h]
    \centering
    \resizebox{\linewidth}{!}{%
    \begin{tikzpicture}[xscale=1.0, yscale=1]
      \draw[->] (0,0) -- (9,0) node[below right] {Time};

      \draw[thick,|-|] (0,1) -- (2,1) node[midway, above] {$t_{RP}$};  
      \draw[thick,|-|] (2,0.5) -- (4,0.5) node[midway, above] {$t_{RCD}$}; 
      \draw[thick,|-|] (2,1) -- (8,1) node[midway, above] {$t_{RAS}$}; 

      \node[below] at (0,0) {PRE};
      \node[below] at (2,0) {ACT};
      \node[below] at (5,0) {RD/WR};
      \node[below] at (8,0) {PRE};

      \draw[dashed,gray] (2,1) -- (2,-0);
      \draw[dashed,gray] (4,1) -- (4,-0);
      \draw[dashed,gray] (8,1) -- (8,-0);
    \end{tikzpicture}
    }%
    \caption{DRAM read/write timing sequence showing $t_{RP}$, $t_{RCD}$, and $t_{RAS}$ intervals.}
    \label{fig:dram_timing}
\end{figure}

Multiple steps must occur during the reading and writing of DRAM. Logically, these include reception of the read/write address, the row decoder selection and activation of the required row, row storage in the row buffer, column decoder selection of requested data, data output from buffer to the memory controller and precharge preparation for next access. Each logical step in the read-write sequence takes a predefined amount of time, and it is essential for each step to occur in the correct order to prevent data corruption/memory failure. Figure~\ref{fig:dram_timing} illustrates the standard timing sequence for a typical DRAM read or write operation. The process begins with a \texttt{PRE} (Precharge) command, which prepares the bitline by precharging it to a stable voltage level (typically $V_{\mathrm{DD}}/2$). This precharge period lasts for $t_{RP}$.

Following the precharge, the \texttt{ACT} (Activate) command is issued to enable access to the desired wordline. This triggers the wordline to rise to a high voltage level, connecting the corresponding cell's storage capacitor to the bitline via the access transistor. The time between the activation and the start of the actual read or write command is known as $t_{RCD}$ (Row to Column Delay), during which charge sharing occurs and the sense amplifier resolves the voltage difference on the bitline. The \texttt{RD/WR} (Read or Write) command is then executed, allowing data to be read from or written to the selected cell. After a minimum duration known as $t_{RAS}$ (Row Active Time), the system returns to the precharge phase to close the row and restore the bitline, readying the DRAM array for the next access cycle.

\section{Parameter Characterization}
\subsection{$R_S$ Measurement under Retention}

To characterize a DRAM cell, we measure the effective leakage resistance $R_S$ under different configurations. The procedure is based on retention profiling: the victim cell is initialized to logic \texttt{1}, refresh is disabled for a sequence of intervals $\{t_1, t_2, \dots\}$, and the cell is read out to detect the earliest flip time $t_{\text{flip}}$. Given the storage capacitance $C_S$ and the sense amplifier threshold $V_{\text{flip}}$, the effective resistance is computed as
\begin{equation}
R_S(N) = \frac{t_{\text{flip}}}{C_S \cdot \ln \!\left( \frac{V_{DD}}{V_{\text{flip}}} \right)},
\end{equation}
where $N$ denotes the neighbor pattern applied to the eight adjacent cells. The baseline resistance $R_S(\text{all-1})$ is obtained when all neighbors are written with logic ‘1’. The complete procedure is summarized in Algorithm~\ref{alg:measure_rs_n}.

\begin{algorithm}[t]
\caption{Measurement of $R_S$ and $R_S(N)$ with Pattern Noise}
\label{alg:measure_rs_n}
\begin{algorithmic}[1]
\State \textbf{Input:} Victim cell $v$, neighbor patterns $\{N_0=\text{all-1}, N_1, \dots, N_k\}$, time intervals $\{t_1,\dots,t_n\}$
\State \textbf{Output:}  $R_S$ and $R_S(N)$ with noise
\For{each pattern $N$}
  \State Write victim cell $v \leftarrow \texttt{1}$
  \State Program neighbors of $v$ according to $N$
  \For{each $t_i$}
     \State Disable refresh for duration $t_i$
     \State Read cell $v$
     \If{read result == \texttt{0}}
        \State $t_{\text{flip}} \gets t_i$, \textbf{break}
     \EndIf
  \EndFor
  \State Compute $R_S(N) = \tfrac{t_{\text{flip}}}{C_S \cdot \ln \!\big( V_{DD}/V_{\text{flip}} \big)}$
\EndFor
\State \Return $\{R_S(N)\}$
\end{algorithmic}
\end{algorithm}

\subsection{Measurement of $R_{S}$ and $R_B$ under Read Disturbance}
To quantify the effect of subthreshold conduction under read disturb stress
(Rowhammer/Rowpress), we adopt a two-pattern methodology that separates the
baseline leakage path from its pattern-dependent extension. In the
\texttt{111} configuration (victim and aggressors set to logic~\texttt{1}),
the bitline is restored to $V_{DD}$ after each read, leaving essentially no
DC bias across the access transistor. Although wordline coupling may slightly
perturb the victim, the absence of a driving potential suppresses additional
subthreshold current. Let $HC_{111}$ be the minimal hammer count that produces
the target number of flips (e.g., 5000) at an effective read disturb rate
$f_{\mathrm{RD}}$; the corresponding flip time is
$T_{111} = HC_{111}/f_{\mathrm{RD}}$, and the baseline effective resistance is
\begin{equation}
  R_{\mathrm{eff}} \;\approx\; R_S^{\mathrm{RD}}
  \;=\; \frac{T_{111}}{C_S \,\ln\!\big(\tfrac{V_{DD}}{V_{\mathrm{flip}}}\big)}.
\end{equation}

In contrast, the \texttt{010} configuration (victim at logic~\texttt{1},
aggressors at logic~\texttt{0}) establishes a non-zero bias between the storage
node and bitline, biasing the access transistor into the subthreshold region
and enabling an additional current component. Let $HC_{010}$ be the hammer
count needed to reach the same flip target, and
$T_{010} = HC_{010}/f_{\mathrm{RD}}$ the corresponding flip time. Defining
\begin{equation}
  E \;\triangleq\; \Big(\tfrac{V_{DD}}{V_{\mathrm{flip}}}\Big)^{T_{010}/T_{111}} ,
\end{equation}
and modelling the discharge as a combination of baseline leakage and a constant
subthreshold term, we obtain the closed-form estimate of the subthreshold
factor
\begin{equation}
  A \;=\; \frac{1}{R_{\mathrm{eff}}}\,
          \frac{V_{DD}-E\,V_{\mathrm{flip}}}{E-1},
\end{equation}
which depends only on measured hammer counts $(HC_{111},HC_{010})$ and known
device parameters $(f_{\mathrm{RD}},C_S,V_{DD},V_{\mathrm{flip}})$. The
parameter $A$ has units of current and quantifies the extra leakage due to
subthreshold conduction; in our model, it directly leads to the bitline-side
effective resistance $R_B^{\mathrm{RD}} \approx V_{TH}/A$.

\section*{Open Science}
To support reproducibility and foster further research, we make all artifacts of this work publicly available at \href{https://osf.io/9gdyw/?view_only=eb552500a9f445d8a56611dc7ea36d25}{Open Science Framework Anonymous Link}.
This includes our FPGA-based measurement framework, implementation of the proposed models,
and security evaluation scripts. 

\section*{Ethical Considerations}
Our experiments were performed on commodity DRAM modules in controlled lab settings, without involving production systems, user data, or human subjects. The attacks are evaluated solely for research purposes. The goal is to inform the community and vendors about security implications to enable more secure memory designs.

\end{document}